\documentclass[aps,a4paper,amsmath,amssymb,twocolumn,
floatfix,groupedaddress]{revtex4-1}
\usepackage[T1]{fontenc}
\usepackage[utf8]{inputenc}

\usepackage{bm,amsmath,amssymb,amsfonts}
\usepackage[dvips]{graphicx}
\usepackage{slashed}
\usepackage{relsize}
\usepackage{braket}
\usepackage{color}
\definecolor{dblue}{rgb}{0.0,0.0,0.7}
\definecolor{dred}{rgb}{0.9,0.0,0.0}
\definecolor{dpink}{rgb}{0.85,0.067,0.49}
\usepackage{hyperref}

\hypersetup{
    pdfnewwindow=true,      
    colorlinks=true,       
    linkcolor=dpink,          
    citecolor=blue,        
    filecolor=blue,      
    urlcolor=blue        
}
\newcommand\ba{\begin{eqnarray}}
\newcommand\ea{\end{eqnarray}}
\newcommand\be{\begin{equation}}
\newcommand\ee{\end{equation}}

\newcommand{\bi}{\bibitem}

\begin{document}

\title{Periodic and aperiodic dynamics of flat bands in diamond-octagon lattice}

\author{Tanay Nag$^1$ and Atanu Rajak$^2$} 
\email{$^1$tnag@sissa.it}
\email{$^2$raj.atanu009@gmail.com}
\affiliation{$^1$ SISSA, via Bonomea 265, 34136 Trieste, Italy }
\affiliation{$^2$ Presidency University, 86/1, College Street, Kolkata 700073, India}

\date{\today}

\begin{abstract}

We drive periodically a two-dimensional diamond-octagon lattice model by switching between two Hamiltonians
corresponding two different magnetic flux piercing through diamond plaquette to investigate the generation of
topological flat bands. We show that in this way, the flatness and topological nature of all the bands of the model
can be tuned and Floquet quasi-bands can be made topologically flat while its static counterpart does not support \textcolor{black}{
the topological band and flat band together. We define here a measure of flatness of quasi-energy bands from their microscopic details that has been justified using the numerical calculations of the Floquet density of states. In the context of Floquet dynamics, we indeed have a better control on the engineering of flat bands.}
Interestingly, we find the generation of flux current due to periodic drives. We systematically analyze the work done
and flux current in the asymptotic limit as a function of system's parameters to show that topology and flatness both share
a close connection to the flux current and work done, respectively. We finally extend our investigation to the aperiodic
array of step Hamiltonian, where we find that the heating up problem can be significantly reduced if the initial state
is substantially flat as the initial large degeneracy of states prevents the system from absorbing energy easily from
the aperiodic driving. In addition, we show that the heating can be reduced if the values of the magnetic flux in the
step Hamiltonians are made small and, also the duration of these fluxes become unequal. Finally, we successfully
explain our finding by plausible analytical arguments.

\end{abstract}

\maketitle

\section{Introduction}

 Tight-binding translationally invariant models 
with local symmetries can exhibit flat bands (FBs) that have   
received a lot of research attention in recent times \cite{das-sharma-prl2011, tang-prl2011, titus-prl2011, pollmann-prb2015, flach-prb2013, 
flach-epl2014, flach-prl2014, flach-prb2015, flach-prl2016, flach-prb2017, vicencio-pra2017, 
denz-apl2017, ajith-prb2017, ajith-prb2018, biplab-prb2018}. These 
FBs, originated from the destructive interference of electron 
hopping, have vanishingly small band-width,
and they host macroscopic number of 
degenerate single particle states. A perturbation  that can lift the degeneracy
thus be able to probe the strongly correlated nature of the eigenstates.
FB can also appear in continuum e.g., Landau levels are 
formed in 2D electron gas in presence of magnetic field. We note that 
completely (partially) filled  Landau levels exhibit integer (fractional)
quantum Hall effect \cite{klitzing-prl1980,laughlin-prb1981}. 
The non-trivial FBs not only
bears a deep connection with the topology \cite{das-sharma-prl2011, tang-prl2011}
but also leads to other intriguing phenomena in condensed matter physics
\cite{tasaki-prl1992, tanaka-prl2003, bitan-prx2014,kauppila-prb2016,
peotta-nc2015,goda-prl2006, shukla-prb2010}.  The experimental 
search has already began in this area of research: FBs have been 
observed in photonic waveguide networks~\cite{vicencio-njp2014,
vicencio-prl2015, mukherjee-prl2015, 
mukherjee-ol2015, longhi-ol2014, xia-ol2016, zong-oe2016,
weimann-ol2016}, exciton-polariton condensates~\cite{masumoto-njp2012, baboux-prl2016}, and 
ultracold atomic condensates~\cite{jo-prl2012, taie-sa2015}.
On the other hand, FBs can be observed in  tight-binding lattice models for
a variety of lattice geometries such as
Lieb~\cite{manninen-pra2010, goldman-pra2011}, 
kagome~\cite{altman-prb2010}, honeycomb~\cite{wu-prl2007}, square~\cite{monika-np2015}, which
can be realized using ultracold fermionic or bosonic atoms in optical lattices. 
\textcolor{black}{ Apart from these, various  non-interacting systems are  found
to exhibit FBs~ \cite{Montambaux18,Jiang19,Lim20,Tang11}. 
Very recently, FBs receive a lot of attention in the context of 
 twisted bi-layer graphene where Coulomb interaction plays a vital role ~\cite{Koshino18,Xian19,Klebl19,balents20}. Surprisingly,  the
 FBs are analyzed in Creutz model in both presence or absence of interactions~\cite{Kuno20}. The existence of FBs in Moir\'e structure is also not restricted to interacting cases only ~\cite{pnas11,haddadi2020}. }

Quite importantly, the 
study of non-equilibrium dynamics of closed quantum systems 
is another growing field of research from theoretical \cite{calabrese06,
rigol08,oka09,kitagawa10,lindner11,bermudez09,patel13,thakurathi13,pal10,nandkishore15,heyl13,budich15,sharma16} as well as 
experimental \cite{bloch-rmp2008,lewenstein12,jotzu14,greiner02,
kinoshita06,gring12,cheneau12,fausti11,rechtsman13} point of view.
In particular, a periodically driven system, with 
Hamiltonian $H(t)=H(t+T)$, $T$ being the period of drive,
yields a non-trivial state of matter while its equilibrium counter-part supports  the trivial 
state. We would
like to mention a few interesting consequences of the periodic drive
in the context of defect and residual energy generation \cite{mukherjee08,russomanno12},
dynamical freezing \cite{das10}, many-body energy localization 
\cite{alessio13}, dynamical localization\cite{nag14,agarwala16},
and quantum information studies \cite{nag16,russomanno16}. 
Interestingly, light induced Floquet graphene \cite{oka09,kitagawa10},
topological insulator \cite{lindner11}, Floquet higher order 
topological phases \cite{nag19} and dynamical generation of edge Majorana \cite{thakurathi13}
are a few examples of dynamical topological phases due to periodic drive.
For the aperiodic drive the system is expected to absorb the 
energy indefinitely unlike the periodic case where non-equilibrium steady state is observed \cite{bhattacharya18,maity18}. 
One can contrastingly show that for a periodically driven non-integrable system,
heating up is most likely to be unavoidable \cite{alessio14}. Although recent studies showed 
that there are specific situations when the heating can be reduced or suppressed~\cite{choudhury14,citro15,abanin15,rajak18,rajak19}.
\textcolor{black}{Interestingly, the aperiodic system also falls into a different class of 
geometrical generalised Gibbs ensemble  \cite{nandy17} periodic  system lies in the  periodic Gibbs ensemble
class \cite{lazarides14}.}
A system with quasi-periodic drive is also studied in the context of topology \cite{Crowley19}.

Now turning into the physics of FBs, 
it is noteworthy that  
nontrivial topology, finite-range hopping and exactly FBs have some interesting interplay between 
them. It has been shown that these three above criteria can not be 
simultaneously satisfied, only two of these  can be realized simultaneously~\cite{chen-jpa2014, read-prb2017}. 
 The spectral flattening technique can adiabatically transform the original Hamiltonian to 
a new one with FB states; however, in this case the  underlying Hamiltonian might be accompanied with the long-range hopping \cite{das-sharma-prl2011}.
The short-range hopping can also be obtained  following some other  optimization techniques~\cite{ronny-prb2016, ronny-prb2017}.

In parallel, the generation of flat bands under periodic drive is also a timely 
research topic. Using periodic drive, one can generate flat bands that do not have any static 
analog. In some cases, these flat bands can show non-trivial topology, i.e., those have non-zero Chern number. 
It has been shown that topological FBs  can emerge not only at zero quasi-energy but also at  $\pm\pi$ quasi-energy   
between two inequivalent touching band points with  opposite  Berry  phase
in a time-periodically driven uniaxial strained graphene nanoribbons \cite{Taboada17}.
Starting from a trivial phase of $s$-wave superconductor, suitably activating or generating the chiral symmetry via 
Floquet dynamics one can obtain Majorana FBs \cite{Poudel15}. Recently, it has been shown that in the presence of an 
external magnetic flux piercing through the diamond plaquettes in the  2D diamond-octagon lattice model 
with short-ranged hopping can support topological FBs \cite{Pal18}. The application of magnetic field breaks 
the time-reversal symmetry of the system. In absence of time-reversal symmetry breaking, the system supports 
FBs, but those are non-topological and not well separated from the other dispersive 
bands. By applying magnetic flux, the flat bands maintain a gap from the other dispersive bands, however,
they show topological 
behavior for some specific choices of magnetic flux.

\textcolor{black}{ 
Given the fact that FBs are observed in non-interacting models \cite{manninen-pra2010, goldman-pra2011,Montambaux18,Jiang19,Lim20,Tang11,Kuno20}, we consider a simple four band model such as two-dimensional diamond-octagon tight-binding lattice model \cite{Pal18}. The photonic
waveguides and optical lattice systems are found to be extremely useful to simulate the lattice structure in lab  \cite{Vicencio15,Mukherjee15_2,jo12,Aidelsburger11}.
Moreover, recent advancement on experimental side in realizing Floquet dynamics allows us to consider the step like driving protocol \cite{exp1,exp2,exp3,exp4,exp5,exp6}.   
We here focus on such a dynamical scheme to optimize the occurrence of topological FBs in the above model.
Considering the equilibrium study on the topological FBs in  presence of flux in the above mentioned lattice model}, our aim is twofold: 1) can one 
generate topological FBs using periodic drive by switching between 
two flux Hamiltonian,
while the underlying static flux Hamiltonian does not support FBs?
2) what is fate of energy absorption under aperiodic drive for these type of systems with FBs? 
\textcolor{black}{Having redefined flatness to quantitatively describe the 
quality of the Floquet FBs, we show that under the variation of the 
temporal width of the two step Hamiltonian and their associated fluxes, 
one can tune topology along with flatness.} \textcolor{black}{Our investigation suggests that 
flux current can be dynamically induced.} We also indicate the connection of topology and flatness to the  flux current and 
excess energy (also referred as residual energy or work done), respectively.
We extend our analysis for 
aperiodic case where the rate of absorption can be regulated  significantly 
with the above set of parameters.  The flux term allows us to study the excess 
energy with initial states having different flatness; most interestingly, we find that for aperiodic driving,
FBs are found to be a better absorber of energy than dispersive band. 
We explain our numerical results with plausible physical 
arguments.

The paper is organized as follows. We describe the lattice model and the dynamical protocol
for the periodic and aperiodic case in Sec.~\ref{model}. There 
we also present the  definition of  the flatness and Chern number for a generic periodically driven system. 
Next, in Sec.~\ref{pd} and Sec.~\ref{apd}, we discuss our main results
following the periodic and aperiodic driving, respectively. We show the variation of flatness and Chern number as a function of the 
driving parameters. In addition, we also study there the stroboscopic behavior of flux current and excess energy. 
We repeat these analysis for the aperiodic case. 
Finally, in Sec.~\ref{summary}, we conclude our work.

\section{The model and the dynamical protocol}
\label{model}
\begin{figure}[ht]
\includegraphics[width=1.0\columnwidth]{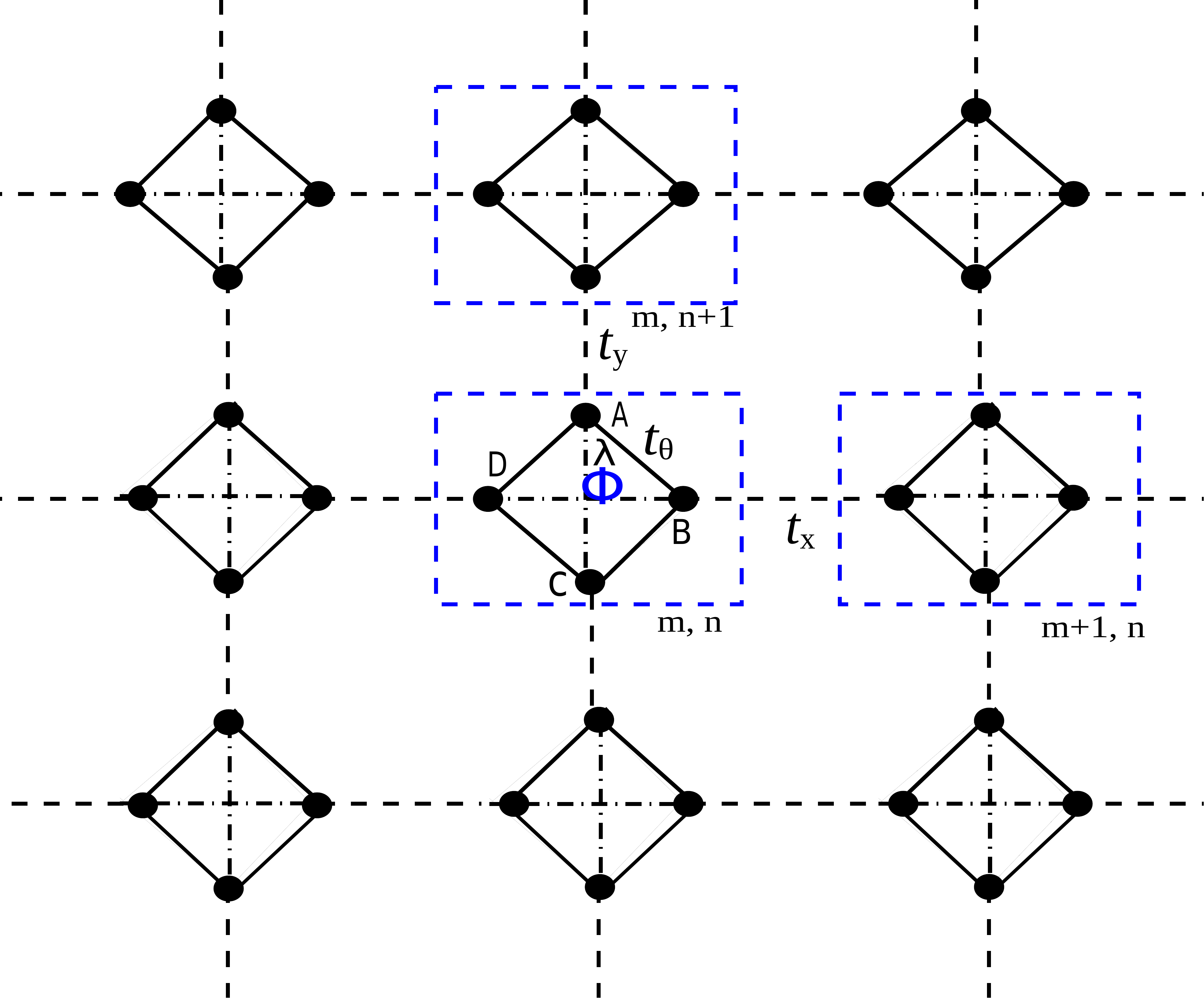}
\caption{ (Color online) The lattice geometry of 2D diamond-octagon lattice is shown. The diamond plaquette, consisting four sites at the vortices $A$, $B$, $C$ and $D$, is depicted by dashed blue boxes that refers to the unit cell. The flux $\phi$ is enclosed by the diamond plaquette. The  electrons while intra-cell hopping along the closed loop through the diamond arms $A\to B\to C \to D \to A$ thus acquire a phase and the these hoppings are denoted by $t_{\theta}$.  The diagonal hoppings  $A\to C$ and $B \to D$, denoted by $\lambda$, do not acquire any phase as they do not enclose a closed loop inside the diamond plaquette. $t_x$ and $t_y$ respectively denote the horizontal and vertical inter-cell hopping between two adjacent diamond plaquettes. 
}
\label{fig:lattice}
\end{figure}

We consider a two dimensional diamond lattice model  where four atomic sites are at the four vortices of the unit cell.
 The basic unit cells, comprising the diamond-shaped loop, are repeated 
periodically in $x$ and $y$ directions to obtain the whole lattice structure.
We consider a uniform magnetic flux perpendicular to the plane of lattice piercing through
the diamond plaquette (i.e., intra-cell flux); this introduces an Aharonov-Bohm phase to the 
hopping parameter when an electron hops along the boundary of a diamond loop. 
We note that there is no inter-cell flux involved. \textcolor{black}{
The lattice structure is depicted in Fig.~\ref{fig:lattice}}.
The tight-binding Hamiltonian 
of this model in Wannier basis can be written as,
\begin{equation}
\bm{H} = \sum_{m,n} \Bigg[ \sum_{i}\epsilon_{i} c_{m,n,i}^{\dagger}c_{m,n,i} + 
 \sum_{i,j} \Big( \mathcal{T}_{ij} c_{m,n,i}^{\dagger}c_{m,n,j} + \textrm{H.c.} \Big) \Bigg ],
\label{eq:hamil-wannier}
\end{equation}
where the first summation runs over the unit cell index $(m,n)$. 
$c_{m,n,i}^{\dagger}$ $(c_{m,n,i})$ is the creation (annihilation) operator for an electron at site $i$ in the 
$(m,n)$-th unit cell and $\epsilon_{i}$ is the on-site potential for the $i$-th atomic site. Here, the diamond plaquette is referred as the unit cell.
The parameter $\mathcal{T}_{ij}$ 
is the hopping strength between the $i$-th and the $j$-th sites, and it can take two possible values depending 
on the position of the sites $i$ and $j$. We denote
$\mathcal{T}_{ij} = t_{x}$ ($t_y$) for an electron hopping between two adjacent diamond plaquettes along 
$x$ ($y$) directions. In addition of the above inter-cell hopping, the model also contains two types of intra-cell hopping.  $\mathcal{T}_{ij} = \lambda$ when an electron hops along the diagonals inside a diamond plaquette. 
On top of that we have another hopping $t_{\theta}$ when 
the electron hops around the  closed loop in a diamond plaquette.
Each diamond plaquette is pierced by an external magnetic flux $\phi$ which incorporates an Aharonov-Bohm 
phase factor to hopping parameter $t_{\theta} \rightarrow t_{\theta} \exp{(\pm i\theta)}$.
Here, $\theta =  \pi \phi / 2 \phi_{0}$, $\phi_{0}=hc/e$ being the 
fundamental flux quantum, the sign $\pm$ in the exponent indicates the direction of the forward and the 
backward hoppings and $\phi$ would be in terms of $\phi_0$. For the rest of the paper, we shall refer $\theta$ as the flux for simplicity. \textcolor{black}{For completeness, we note that one can consider an inter-cell flux enclosed by the octagon plaquette. However, the properties of the system might change once the nature of the flux changes. For example, the octagon flux directly incorporates the diamond  flux in addition to the inter-cell fluxes.}
\textcolor{black}{We restrict ourselves to non-interacting case where single-particle state can describe the essential physics of flat bands. In order to incorporate more realistic effects such as correlated phenomena, one has to consider interacting version of the model.  We comment on the possible effects of interactions in our model at the end of our paper (see \cite{supple} for detail).}

Using discrete Fourier transform, the momentum space description of the Hamiltonian in 
Eq.~\eqref{eq:hamil-wannier} can be read as,
\begin{equation}
\bm{H} = \sum_{\bm{k}} {\Psi}^{\dagger}_{\bm{k}} {\mathcal{H}}_{\bm k}(\theta) {\Psi}_{\bm{k}},
\label{eq:hamil-mom} 
\end{equation}
where $
{\Psi}^{\dagger}_{\bm{k}} \equiv
\left(\begin{matrix}
c^{\dagger}_{k_x,k_y,A}  &  c^{\dagger}_{k_x,k_y,B}  
&  c^{\dagger}_{k_x,k_y,C} & c^{\dagger}_{k_x,k_y,D} 
\end{matrix}\right)$,
and ${\mathcal{H}}_{\bm{k}} (\theta)$ is given by,
\begin{align}
{\mathcal{H}}_{\bm k}(\theta) =
\left(\def\arraystretch{1.5} \begin{matrix}
0  &  t_{\theta}e^{i\theta}  &  t_y e^{ik_y}+\lambda  &  t_{\theta} e^{-i\theta} \\
t_{\theta}e^{-i\theta}  &  0  &  t_{\theta}e^{i\theta}  &  t_x e^{ik_x}+\lambda \\
t_y e^{-ik_y}+\lambda  &  t_{\theta} e^{-i\theta}  &  0  &  t_{\theta} e^{i\theta} \\
t_{\theta} e^{i\theta}  &  t_x e^{-ik_x}+\lambda  &  t_{\theta} e^{-i\theta}  &  0 \\ 
\end{matrix}\right).
\label{eq:ham}
\end{align}
We have taken $\epsilon_{i}=0$, $i\in \{A,B,C,D\}$ being the 4 vortices of the diamond plaquette. One can extract all the interesting features about 
the band structure of the static system as well as driven system  as presented in the next section.

Now we shall describe the dynamical scheme. In our present work, we study the effect of a generic aperiodic 
temporal variation of the flux Hamiltonian ${\mathcal{H}}_{\bm{k}} (\theta)$. We  consider the step-like driving protocols 
with a binary disorder in the amplitude of driving. Interestingly, one can easily recover the periodic limit from the above driving protocol
without any loss of 
generality. We consider two Hamiltonians ${\mathcal{H}}_{\bm k}(\theta_0)$ and ${\mathcal{H}}_{\bm k}(\theta_1)$ 
in one time-period having magnetic flux $\theta_0$ and $\theta_1$, respectively.
The explicit driving protocol is given by 
\begin{align}\label{eq:3}
{\mathcal{H}}_{\bm k}(t) & = {\mathcal{H}}_{\bm k}(\theta_0) ~~~\quad  \text{for}~~ (n-1)T< t <n' T,   && \\ \nonumber
& = (1-g_n){\mathcal{H}}_{\bm k}(\theta_0) + g_n {\mathcal{H}}_{\bm k}(\theta_1) \quad\text{for} ~~ n' T< t <nT, \quad 
\end{align} 
with $n'=n-1+\alpha$. Here, $T$ is the time period, $\alpha$ is a parameter defined between $0$ and $1$ and 
$n$ refers to the $n$-th stroboscopic period.
As a result, the system is evolved by the Hamiltonian $\mathcal{H}_{\bm k}(\theta_0)$ for time duration $\alpha T$, 
whereas the Hamiltonian $\mathcal{H}_{\bm k}(\theta_1)$ or $\mathcal{H}_{\bm k}(\theta_0)$ can take care the dynamics for the remaining time 
$(1-\alpha)T$ within  one time-period depending on the value of $g_n$.
The two step Hamiltonians in $k$-space refer to the system on diamond-octagon lattice in real space with
two different fluxes $\theta_0$ and $\theta_1$, respectively.
The random variable $g_n$ takes the value either $1$ with
probability $p$ or $0$ with probability $(1-p)$ chosen  from a  Binomial distribution. 
For $g_n=0$, the system evolves with the Hamiltonian having flux $\theta_0$ in the $n$-th time
period within the time interval  $(n-1)T$ to $nT$, while 
$g_n=1$ corresponds to the step driving i.e., 
the subsequent evolution is governed by ${\mathcal{H}}_{\bm k}(\theta_0)$ and ${\mathcal{H}}_{\bm k}(\theta_1)$ 
for time duration $\alpha T$ and $(1-\alpha)T$, respectively, within a time period.

We note that 
if initial state is chosen to be the ground state of 
$ {\mathcal{H}}_{\bm k}(\theta_0)$, the case with $g_n=0$ refers to the free evolution,
whereas, for $g_n=1$, the dynamics is non-trivial. Thus the random variable 
$g_n$ introduces aperiodicity in the system. 
In addition, the parameter $\alpha$ determines time duration of switching between two Hamiltonians ${\mathcal{H}}_{\bm k}(\theta_0)$ 
and ${\mathcal{H}}_{\bm k}(\theta_1)$, over one time-period in the case of $g_n=1$.
As a result, for $g_n=1$ and $\alpha=0.5$ (i.e., equally divided time duration within a single time-period),
the dynamics of the initial state is governed by the product 
of two unitary matrices (see Eq.~(\ref{eq_FO})) which is designated by Bang-Bang protocol. 
For $p=1$ i.e., $g_n=1$ for all $n$, the problem reduces to periodic one which can be formulated using the Floquet theory.
The details are given below.

In general, the initial state  $|\Psi_{\bm k}(\theta_{\rm ini},0)\rangle$ can be considered as the ground state of the
Hamiltonian ${\mathcal H}_{\bm k}(\theta_{\rm ini},t=0)$ with flux $\theta_{\rm ini}$.
One can choose $\theta_{\rm ini}=\theta_0$ for simple situation where the dynamics starts from the ground state of the first step 
Hamiltonian, otherwise, for  $\theta_{\rm ini} \ne \theta_0$, it is always a non-eigenstate evolution even for $g_n=0$. 
 In this way, we have a 
complete freedom on the choice of initial state for the subsequent dynamics.
For the periodic driving in Sec.~\ref{pd}, we restrict ourselves to the case $\theta_{\rm  ini}=\theta_0$ while for 
aperiodic driving in Sec.~\ref{apd}, $\theta_{\rm  ini}\ne \theta_0$ and  $\theta_{\rm  ini}=\theta_0$ 
both the situation are considered.  We shall also explore the situation where 
$\theta_0\neq 0$ and $\theta_0 \neq \theta_1$.
\textcolor{black}{Throughout this paper we have considered $\hbar=1$. The frequency $\omega$ is the dimension 
of inverse of time. It enters in the system through the Floquet operator where it appears with the multiplication 
of energy, and the whole quantity becomes dimensionless~(see around Eq.~(\ref{eq_ft4})). 
Similarly, $t_x$, $t_y$ and $\lambda$ are dimension of energy. We have assumed here $t_x$,
$t_y$ and $\lambda$ to be $1$ such that all the  relevant observables are measured in units of these hopping strengths (see Sec.~\ref{expt} for experimental estimations). In this work, we consider the high frequency limit for both periodic and aperiodic drives so that our results are valid well above the resonance limit \cite{russomanno12,nag14,bhattacharya18}. 
This limit is set by considering the driving frequency greater than the band-width for the equilibrium case and assumed 
to be $\omega=8$ for all our numerical calculations.}
We average over $10^3$ realization for the aperiodic case.

Coming back to Floquet theory, we here consider a time periodic Hamiltonian
$H(T+t)=H(t)$ where $T$ being the time period. In our case, 
for each ${\bm k}$ mode, we then have
${\mathcal{H}}_{{\bm k}}(t+T)={\mathcal{H}}_{{\bm k}}(t)$. Using the Floquet formalism, one can define a Floquet evolution operator 
 ${\mathcal{F}}_{\bm k}(T) = {\mathcal T} \exp \left( -i \int_{0}^{T}{\mathcal{H}}_{{\bm k}}(t) dt \right)$, where
 ${\mathcal T}$ denotes the time ordering
 operator.  One can caste the Floquet operator using quasi-energy 
 $\mu^{(j)}_{\bm k}$  and quasi-states $|\Phi^{(j)}_{\bm k}(T)\rangle$
 ${\mathcal F}_{\bm k}(T)=\sum_{j} e^{-i\mu^{(j)}_{\bm k} T}
 |\Phi^{(j)}_{\bm k}(T)\rangle \langle 
 \Phi^{(j)}_{\bm k}(0)|$. The point to note here is that $|\Phi^{(j)}_{\bm k}(t+T)\rangle=
 |\Phi^{(j)}_{\bm k}(t)\rangle$ for the time periodic Hamiltonian. 
Therefore,  an arbitrary initial  state  $|\Psi_{\bm k}(t=0)\rangle$ 
can be decomposed in the Floquet basic: 
$|\Psi_{\bm k}(0)\rangle= \sum_{j} r^{(j)}_{\bm k} |\Phi^{(j)}_{\bm k}(0)\rangle$, where 
$r^{(j)}_{\bm k}= \langle\Phi^{(j)}_{\bm k}(0)|\Psi_{\bm k}(0)\rangle$
is the overlap of the Floquet modes and the initial 
wave-function. Combining the above two relations, we get the time evolved wave-function at $t= nT$
\be
|\Psi_{\bm k} (nT)\rangle={\mathcal{F}}_{\bm k}(nT)|\Psi_{\bm k}(0)\rangle=
\sum_{j} r^{(j)}_{\bm k} e^{-i\mu^{(j)}_{\bm k} nT} |\Phi^{(j)}_{\bm k}(T)\rangle.
\label{eq_ft4}
\ee

In our case of the periodic step driving with Hamiltonian ${\mathcal{H}}_{\bm k}(\theta_0)$
and ${\mathcal{H}}_{\bm k}(\theta_1)$, 
${\mathcal{F}}_k$ can be exactly written, 
in the form of,
\begin{equation}
{\mathcal{F}}_{\bm k}(T) = \exp(-i  {\mathcal{H}}_{\bm k}(\theta_1) (1-\alpha) T)
\exp(-i \alpha T  {\mathcal{H}}_{\bm k}(\theta_0) ).
\label{eq_FO}
\end{equation}
Turning into the aperiodic case $0 < p < 1$, there exists
a probability of $(1-p)$ to evolve the system with the Hamiltonian ${\mathcal{H}}_{\bm k}(\theta_0)$ in  every complete period.
We can now express the corresponding evolved state after $n$ complete periods as
\begin{equation}
\lvert \Psi_{\bm k}(nT)\rangle = \mathcal {U}_{\bm k}(g_n) \mathcal{U}_{\bm k}(g_{n-1}). . .  . . . . 
\mathcal{U}_{\bm k}(g_2)  \mathcal{U}_{\bm k}(g_1) \lvert \Psi_{\bm k}(\theta_{\rm ini},0) \rangle
\end{equation}
with the generic evolution operator  given by,
\begin{equation}
\mathcal{U}_{\bm k}(g_n)=\begin{cases}
{\mathcal{F}}_{\bm k}(T), & \text{if $g_n = 1$}.\\
U^{0}_{\bm k}(T), & \text{if $g_n = 0$}.
\end{cases}
\label{eq_protocol}
\end{equation}
where $\boldmath{\cal{F}}_k(T)$ is the usual Floquet operator as given in Eq.~(\ref{eq_FO}). On the 
other hand, $U^0_{\bm k}(T)= \exp(-i {\mathcal H}_{\bm k}(\theta_0) T)$ is the time evolution operator 
using the first step Hamiltonian ${\mathcal H}_{\bm k}(\theta_0)$.

We shall now compute the instantaneous 
stroboscopic energy $e_{\bm k}(nT)$ following both periodic and aperiodic driving. 
We note that at the  stroboscopic  instant $t=nT$, the Hamiltonian ${\mathcal H}_{\bm k}(\theta_0,t=nT)$ 
governs the system; this is the starting Hamiltonian also during the course of dynamics: 
${\mathcal H}_{\bm k}(\theta_0,t=nT)={\mathcal H}_{\bm k}(\theta_0,t=0)$.
For periodic driving $e_{\bm k}(nT)$ is simply given by
$e_{\bm k}(nT)=\langle\Psi_{\bm k}(nT)|{\mathcal H}_{\bm k}(\theta_{0},t=0)|\Psi_{\bm k}(nT)\rangle$. On the 
other hand, for aperiodic driving, instantaneous 
stroboscopic energy becomes 
\begin{eqnarray} 
e_{\bm k}(nT) & = & \langle \Psi_{\bm k}(0) \rvert \mathcal{U}_{\bm k}^{\dagger}(g_1) 
\mathcal{U}_{\bm k}^{\dagger}(g_2) ....... \mathcal{U}_{\bm k}^{\dagger}(g_{n-1}) \mathcal{U}_{\bm k}^{\dagger}(g_n) 
\nonumber \\
&\times & {\mathcal H}_{\bm k}(\theta_0,t=0) \mathcal{U}_{\bm k}(g_n) \mathcal {U}_{\bm k}(g_{n-1})........
\mathcal{U}_{\bm k}(g_2)   \nonumber \\
&& \mathcal{U}_{\bm k}(g_1) \lvert\Psi_{\bm k}(0)\rangle \nonumber \\
\label{eq_rs1}
\end{eqnarray}
Using this, we can calculate the 
 residual energy $W$ (also known as work done and excess energy) in the  driven system  defined as
\begin{equation}
W(nT)=\frac{1}{L^2}\sum_{\bm k}(e_{\bm k}(nT)-e_{\bm k}^{\rm ini}(0)),
\label{eq:re}
\end{equation}
where $e_{\bm k}^{\rm ini}(0)=\langle\Psi_{\bm k}(\theta_{\rm ini},0)|{\mathcal H}_{\bm k}(\theta_{0},t=0)|\Psi_{\bm k}(\theta_{\rm ini},0)\rangle$. 
Similar to the residual energy at finite time, we can calculate it at infinitely long time when the oscillating 
terms averages out to zero. For periodic driving the asymptotic limit of excess energy 
can be written as 
\begin{eqnarray}
W(n\to \infty)&=&\frac{1}{L^2}\sum_{{\bm k}}\Big[\sum_{j=1}^4|r_{\bm k}^j|^2
\langle\Phi_{\bm k}^j|{\mathcal H}_{\bm k}(\theta_{\rm ini},t=0)|\Phi_{\bm k}^j\rangle \nonumber \\
&-& e_{\bm k}^{\rm ini}(0)\Big].
\label{eq:re_infinite}
\end{eqnarray}
\textcolor{black}{ The cross terms carrying the imaginary ``$i$'' inside the exponential: 
$r^p_{\bm k} (r^q_{\bm k})^*\exp(i\mu^p_{\bm k}t-i\mu^q_{\bm k}t)$ with $p\ne q$, do not contribute to the stationary value after momentum summation, as determined by the $|r^p_{\bm k}|^2$, for $n \to \infty$. These 
time-independent terms would eventually survive to yield the non-equilibrium steady state value of the observables~\cite{russomanno12,nag14,bhattacharya18,Lazarides14}.}

We also calculate flux-current $J_{\theta}$ to study the effect of 
Floquet driving in the system having FBs. We define the flux-current operator as $\hat{J_{\theta}}=\frac{\partial {\mathcal H}_{\bm k}(\theta)}{\partial\theta}$, 
given by

\begin{align}
\hat{J_{\theta}} =
\left(\def\arraystretch{1.5} \begin{matrix}
0  &  it_{\theta}e^{i\theta}  &  0  &  -it_{\theta} e^{-i\theta} \\
-it_{\theta}e^{-i\theta}  &  0  &  it_{\theta}e^{i\theta}  &  0 \\
0  &  -it_{\theta} e^{-i\theta}  &  0  &  it_{\theta} e^{i\theta} \\
it_{\theta} e^{i\theta}  &  0  &  -it_{\theta} e^{-i\theta}  &  0 \\ 
\end{matrix}\right).
\label{eq:jtheta}
\end{align}
We note that the flux-current represents the intra-loop current within the diamond unit cell; that is why 
it does not depend on ${\bm k}$. 
Now the current associated with a state $|\Psi_{\bm k}\rangle$ is given by the expectation value of the operator at that state: 
$\langle \hat{J_{\theta}}\rangle= \langle \Psi_{\bm k}|\hat{ J_{\theta}}|\Psi_{\bm k}\rangle$. In a similar spirit,
we can define current along $x$ and $y$ directions. However, it can be 
shown that these current identically vanishes in the ground-state for $\theta_{\rm ini}=0$ referring to the fact that there is no 
inter-loop current present. On the other hand, $\langle \hat{J_{\theta}}\rangle$ remains finite in the 
ground-state only when $\theta_{\rm ini}\ne 0$.

We can determine 
the current of any particular static or Floquet band, and also the total current. The flux current 
in the stroboscopic time evolved state is given by (using Eq.~\eqref{eq_ft4})
\begin{equation}
J_{\theta}(nT)=\frac{1}{L^2}\sum_{{\bm k}}\sum_{j,j'=1}^4r_{\bm k}^j(r_{\bm k}^{j'})^*e^{-i(\mu_k^j-\mu_k^{j'})nT}
\langle\Phi_{\bm k}^j|J_{\theta}|\Phi_{\bm k}^j\rangle.
\label{jtheta_time1}
\end{equation}
At asymptotically long time 
the total $\theta$-current can be expressed as
\begin{equation}
J_{\theta}(n\rightarrow\infty)=\frac{1}{L^2}\sum_{{\bm k}}\sum_{j=1}^4|r_{\bm k}^j|^2
\langle\Phi_{\bm k}^j|J_{\theta}|\Phi_{\bm k}^j\rangle,
\label{jtheta_infty}
\end{equation}
where the oscillating terms of the Eq.~(\ref{jtheta_time1}) will be decayed to zero. \textcolor{black}{The stationary value of flux current is again obtained after summing over the oscillating cross term  as done for Eq.~\ref{eq:re_infinite}}.

We shall now introduce the definition of flatness of the band and quasi-band which will be applicable for static as well as 
time-dependent cases, respectively. Usually the flatness is defined by the 
ratio between band-gap and band-width (see \cite{supple} for details). Now, in this definition, flatness 
can be high once the band-gap $\gg$ band-width, even though the band width is 
significantly large. To overcome this problem, we consider a microscopic definition 
where we calculate the band-width for all points in the BZ and compare it with the 
absolute band gap of the effective one-dimensional system (i.e., $\propto (1/L)$ with system size $L$). 
Therefore, in our alternative definition of flatness, we compare the ratio between the 
local band-width of $i$-th band  $e^i_{\bm k}
-e^i_{\bm k'}$  between ${\bm k}=(k_x,k_y)$ and ${\bm k'}$
to the  absolute gap with a small number $\eta$. \textcolor{black}{ Given the fact that FBs are associated with vanishingly small kinetic energy of the quasi-patricles, we can use the velocity to define the flatness of a given energy band.} 
Now we shall formulate it 
mathematically in detail using the group velocity.  The dispersive nature of 
the energy can be qualitatively
computed using the group velocity for $i$-th band along $x$ and $y$
direction
\be
{v}^{i,x(y)}_{\bm k}=
\frac{e^i_{\bm k} - 
e^i_{\bm k' (k'')} }{\Delta}
\label{eq:flatness_1}
\ee
with ${\bm k'}=(k_x-\Delta k_x,k_y)$,  ${\bm k''}=(k_x,k_y-\Delta k_y)$
and $\Delta =2\pi/L$ is the difference between 
two subsequent ${\bm k}$ points.
We can now calculate the quantity for each point ${\bm k}$
inside the BZ: ${\mathcal V}^i_{\bm k}=
\sqrt{(v^{i,x}_{\bm k})^2 + (v^{i,y}_{\bm k})^2}$.  
We define the flatness from the fraction of points in the BZ for which 
${\mathcal V}^i_{\bm k} < \eta$, $\eta=0.02$. 
\textcolor{black}{We discuss about the choice of $\eta$ to calculate the flatness in Sec.~II of the supplementary material \cite{supple}.}
Let us assume that $l$ is the number of points in the BZ satisfying this criterion, 
flatness is then given by $l/L^2$ such that it is normalized:
$F=l/L^2$. 
This criterion means the magnitude of 
resultant velocity  becomes vanishingly small which is essentially 
reflecting the fact that $i$-th band  is considered to be non-dispersive
once $e^i_{\bm k}
-e^i_{\bm k'} < (2\pi/L) \eta$ for large $L$.

On the other hand, for periodic Floquet driving, 
we use quasi-energy $\mu^{(i)}_{\mathbf k}$ instead of  $e^i_{\bm k}$ to compute the 
stroboscopic flatness. Therefore, quasi-energy band can be contemplated as FB 
if $\mu^{(i)}_{\bm k}-\mu^{(i)}_{\bm k'} < (2\pi/L) \eta$. 
Now for the case of aperiodic dynamics, flatness has to be 
described as a function the number of stroboscopic intervals. This dynamical flatness 
is defined from the instantaneous energy $e_{\bm k}(nT)=
\langle \Psi_{\bm k}(nT) \rvert  {\mathcal{H}}_{\bm k}(\theta_0)
\lvert \Psi_{\bm k}(nT) \rangle$, where $|\Psi_{\bm k}(nT)\rangle$
is the time-evolved wavefunction. Similar to definition of 
static flatness, we here perform the derivative on $e_{\bm k}(nT)$ w.r.t. $k_x$ and 
$k_y$ to compute $v^{x(y)}_{\bm k}(nT)$
\be
v^{x(y)}_{\bm k}(nT)=\frac{e_{\bm k}(nT) - 
e_{\bm k' (k'')}(nT) }{\Delta }
\label{eq:flatness_2}
\ee
We can now calculate the resultant velocity 
 ${\mathcal V}_{\bm k}(nT)$ that would measure of the flatness of the time evolved band.


In order to find whether an energy band is topologically non-trivial, one can calculate Chern number $C$ for that band.
A topological band is characterized by finite non-zero integer value of $C$ while $C=0$ represents the trivial nature. 
For the static system it is calculated using the normalized wave function of $n$-th band, $|n({\bm k})\rangle$ such that 
${\mathcal H}_{\bm k}|n({\bm k})\rangle=E_{n}({\bm k})|n({\bm k})\rangle$. The  Berry curvature of $n$-th band 
using the standard formula~\cite{haldane-prl2004} is given by,
\begin{eqnarray}
&&{\bm \Omega}_{n} (\bm{k}) =
-{\rm Im}\sum_{m\neq n}\frac{\langle n({\bm k})|\nabla\mathcal{H}_{\bm k}|m({\bm k})\rangle\times\langle m({\bm k})|\nabla\mathcal{H}_{\bm k}|n({\bm k})\rangle}{(E_n({\bm k})-E_m({\bm k}))^2}.\nonumber \\
\label{eq:BC}
\end{eqnarray}
Using Eq.~\eqref{eq:BC}, one can easily evaluate the value of the Chern number 
for $n$-th band of the system using the following expression,
\begin{equation}
\textcolor{black}{
C = \dfrac{1}{2\pi}\int_{BZ} \bm{\Omega}_{n} (\bm{k}). d^2{\bm k}},
\label{eq:Chern}
\end{equation} 
where $BZ$ stands for the first Brillouin zone of the corresponding lattice structure.
\textcolor{black}{ We note that the Berry curvature  is a three component vector ${\bm \Omega_n}=(\Omega_n^x,\Omega_n^y,\Omega_n^z)$ while Chern number $C$ is a scalar. 
The area element in two-dimensional momentum space is given by $d^2{\bm k}=dk_x dk_y \hat{z}$. The Chern number measures the Berry flux enclosed by the closed surface.  }

\textcolor{black}{We note that for driven system, the topological characterization is subtle
where Chern number can be found to be insufficient to provide the complete topological description \cite{rudner13,fehske17,kitagawa10,nathan15,fehske18} (see \cite{supple} for detail). However, we restrict ourselves to Chern number only provided the fact that a quasi-static interpretation can work when the system is driven in the  high frequency limit \cite{aoki16}. 
We here  explicitly show how the Chern number of a given band for a static system can be generalized to a driven system.}
In the case of periodic driving,
the Berry curvatures of Floquet bands are obtained by replacing $E_n(\bm k)$ and $|n({\bm k})\rangle$ with $ \mu^{(n)}_{\bm k}$ and 
$ |\Phi^{(n)}_{\bm k}(T)\rangle $, respectively, while the static Hamiltonian is replaced by the time-independent Floquet Hamiltonian. 
In order to compute the Chern number numerically, we use the method suggested in Ref.\cite{fukui14} with $ \mu^{(n)}_{\bm k}$ and 
$ |\Phi^{(n)}_{\bm k}(T)\rangle $. \textcolor{black}{We reiterate that the complete dynamical description of the topological phase requires much more attention. For example,  a dynamic
phase with Chern number $C = 0$ can host edge states \cite{rudner13,nag21}.
This situation can only appear for driven system and does not have any static analogue. In the high frequency limit, the consecutive Floquet Brillouin Zone in the frequnecy space can be found to be well separated allowing us to adopt Chern number description for a given quasi-energy band.}


%
%


\section{Results}
\label{res}

\subsection{Periodic driving}
\label{pd}

\begin{figure}[ht]
\includegraphics[width=1.0\columnwidth]{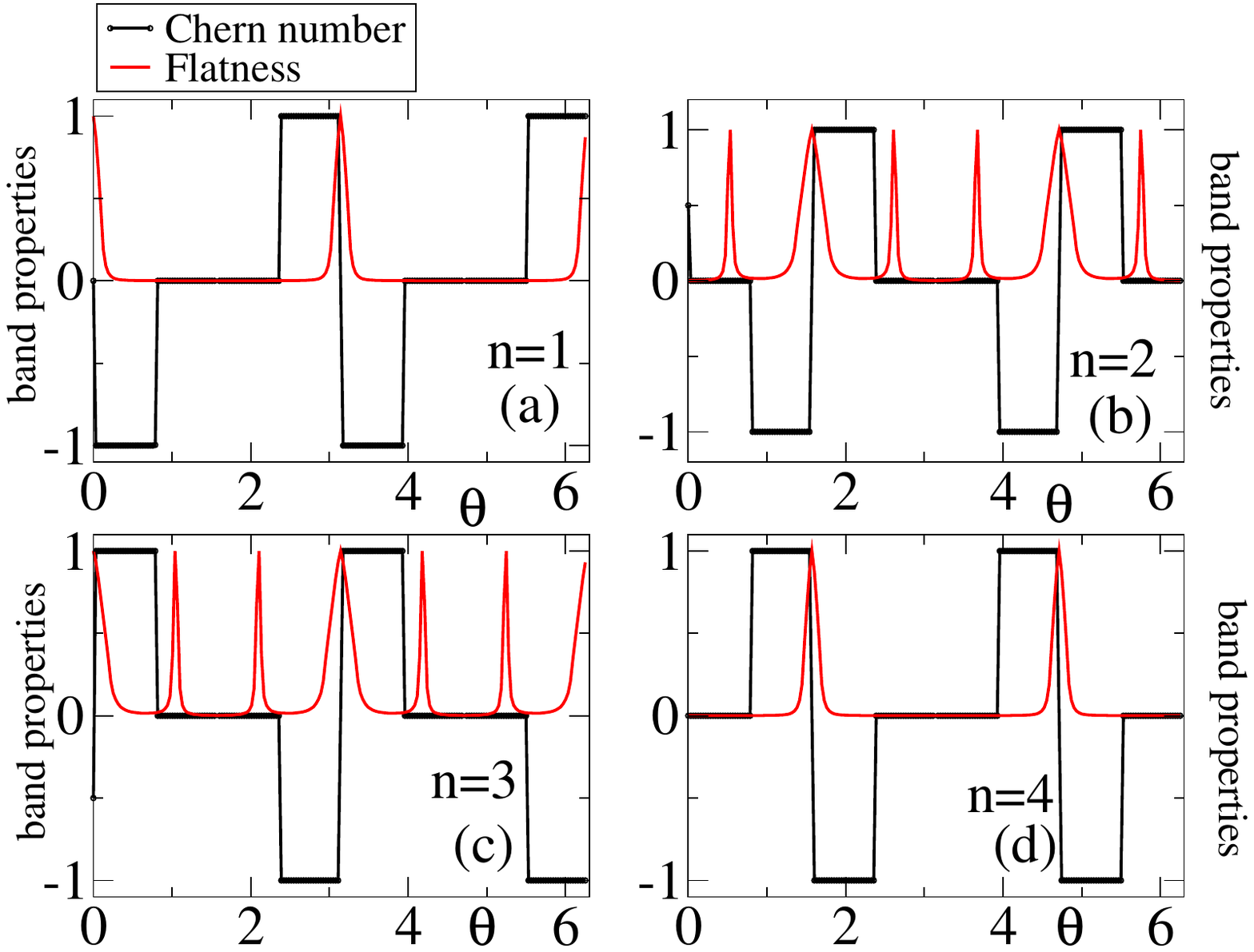}
\caption{ (Color online) Plot depicts the variation of Chern number $C$ and Flatness $F$
as a function of flux $\theta$ for all the energy bands
obtained from the static Hamiltonian (\ref{eq:ham}): $n=1$ in (a),
$n=2$ in (b),  $n=3$ in (c) and  $n=4$ in (d). 
For $n=1,3$, we see that topological flat band is 
maximally probable around $\theta=0$, $\pi$ and $2\pi$. 
While for $n=2,4$, one can observe the topological flat band 
around $\theta=\pi/2$, $3 \pi/2$. 
}
\label{fig:fig1}
\end{figure}

We first study the flatness $F$ and Chern number $C$ for static Hamiltonian ${\mathcal H}_{\bm k}(\theta)$ (\ref{eq:ham})
as shown in Fig.~\ref{fig:fig1}. The static Hamiltonian contains
the magnetic flux term. 
The flatness $F$ of each static energy band $e_{\bm k}^{n=1,4}$ is calculated from the group velocity associated with that energy band. On the other hand, the Chern number $C$ is found from the  eigenstates  $|\Psi_{\bm k}\rangle$ of the static Hamiltonian.
 The bands can simultaneously exhibit
non-trivial topology and high flatness ratio at some specific values of 
$\theta$. For $n=1$ and $3$, topological FBs appear around $\theta =0$, $\pi$ and 
$2\pi$ (see Fig.~\ref{fig:fig1}(a,c)). While for $n=2$ and $4$, topological FBs 
arise around $\theta=\pi/2$ and $3 \pi/2$ (see Fig.~\ref{fig:fig1}(b,d)). 
Therefore, for most of the values of $\theta$, the system remains non-topological, however, 
for $\theta=m\pi/6$ with $m=1,5,7$ and $11$, the $n=2$ band becomes almost flat.
Similarly, trivial FBs appear for $n=3$ at $\theta=m'\pi/3$
with $m'=1,2,4$ and $5$. In a nutshell, all the bands in the static model
support the topological FBs within a very small window of $\theta$.
We also observe that, throughout the whole range of $\theta$, either $(1,3)$-th or $(2,4)$-th energy bands have 
non-trivial topology.
The common feature observed here is that Chern number reverses its sign 
when the flatness becomes maximum except $\theta=m\pi/6$ and $m'\pi/3$ as observed for $n=2$ and $3$, respectively.
Our aim is to manipulate this window of $\theta$ for the existence of topological 
FB along with the reversal of Chern number under Floquet driving. 
\textcolor{black}{The conventional definition of  the flatness is extensively discussed in SI \cite{supple}. }

\begin{figure}[ht]
\includegraphics[clip,width=1.0\columnwidth]{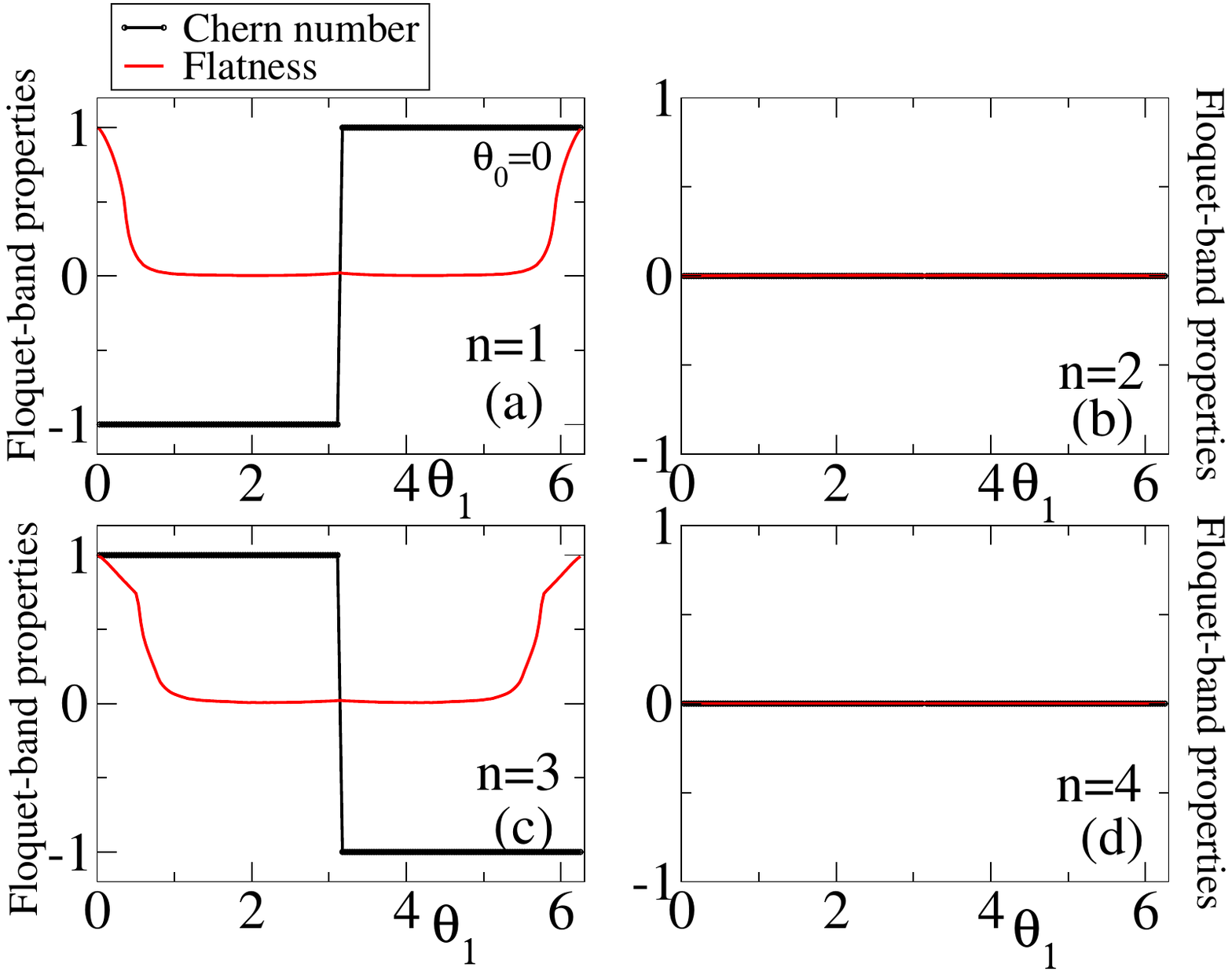}
\caption{ (Color online)
Plot depicts the variation of the Chern number 
$C$ and the flatness $F$
as a function of flux $\theta_1$ for all the bands
obtained from the Floquet operator Eq.~(\ref{eq_FO}): $n=1$ in (a),
$n=2$ in (b),  $n=3$ in (c) and  $n=4$ in (d). 
For $n=1,3$ only, we see that topological flat band is 
 probable around $0<\theta_1<\pi/3$. Here, $\theta_0=0$.
 We note that $\theta_{\rm ini}=\theta_0$ and $\alpha=0.8$.
}
\label{fig:fig2_a}
\end{figure}

\begin{figure}[ht]
\includegraphics[clip,width=1.0\columnwidth]{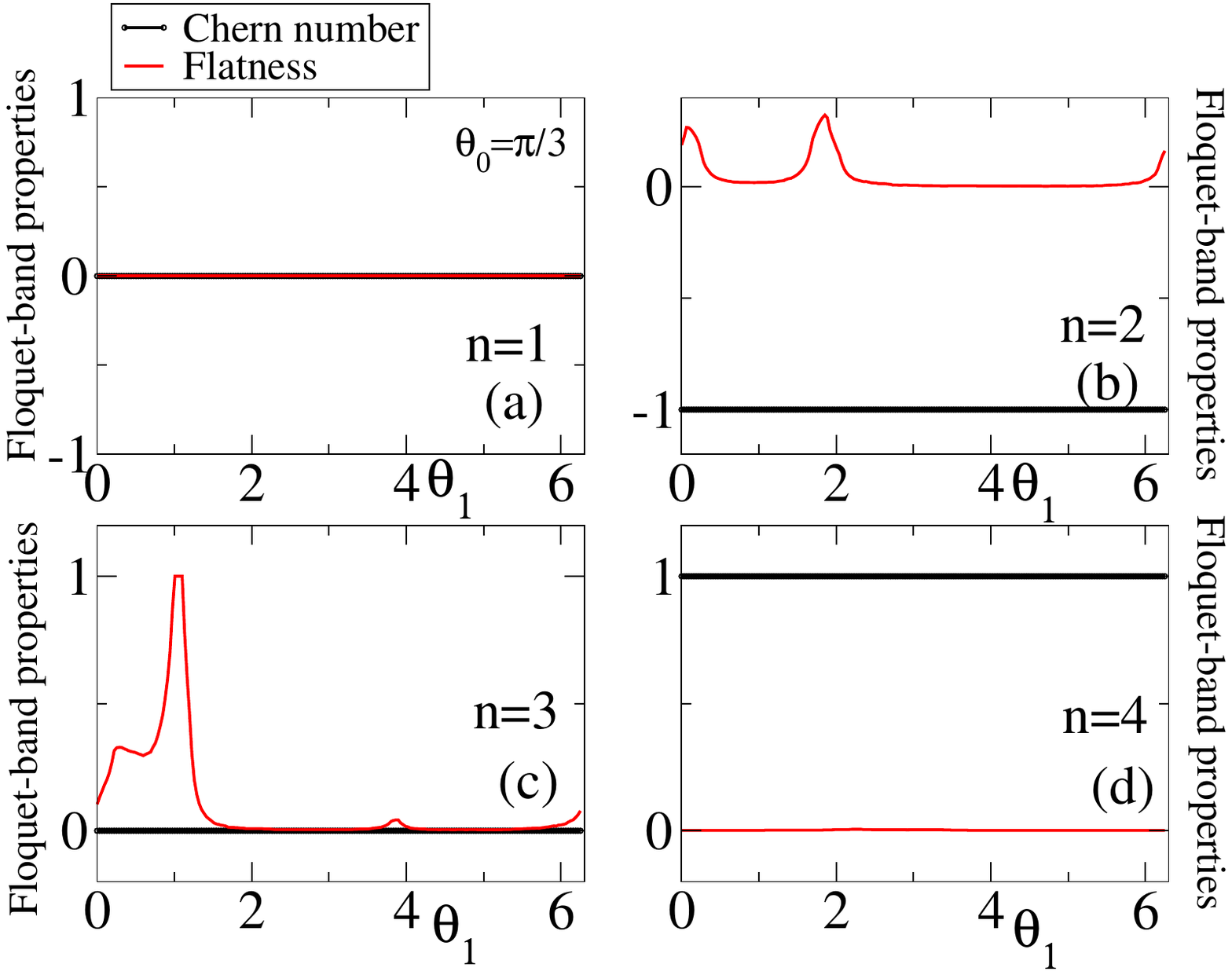}
\caption{ (Color online)
We repeat Fig.~(\ref{fig:fig2_a}) for $\theta_0=\pi/3$.
The topological flat band exists here around $\theta_1=0.1$ and $1.85$
for the band $n=2$ only.
}
\label{fig:fig2_c}
\end{figure}

In order to study the effect of Floquet driving on topology and flatness,
obtained using quasi-states $|\Phi^{(n)}_{\bm k}\rangle$ and quasi-energies $\mu^{(n)}_{\bm k}$, 
we numerically calculate the Chern number and the flatness for the step driving 
with Hamiltonian ${\mathcal{H}}_{\bm k}(\theta_0)$ and ${\mathcal{H}}_{\bm k}(\theta_1)$
as shown in Fig.~\ref{fig:fig2_a}. We consider here $\theta_{\rm ini}=\theta_0=0$.
To find the dependence of $\theta_0$ on the results,
we further repeat our calculation with $\theta_0=\pi/3$ as depicted in 
Fig.~\ref{fig:fig2_c}. 
From Fig.~\ref{fig:fig2_a}, we observe that the topology and the flatness 
are completely suppressed for $n=2$ and $n=4$ Floquet bands with
$\theta_0=0$. As shown in above mentioned figures, in these driving cases, we have another flux parameter 
$\theta_1$ which is tuned to find topology and flatness in quasi-energy bands.
We observe that the region of topological flat band with respect 
to $\theta_1$ increases for $n=1$ and $n=3$ Floquet bands 
as compared to the static case. It is very interesting to note that the value of $\theta_0$ 
is almost same with $\theta_1$ around which the expansion of the flatness is observed.
\textcolor{black}{The important point to note here is that for $\theta_0=\theta_1=\pi/3$,
$n=3$ Floquet band, obtained from  $|\Phi^{(n)}_{\bm k}(T)\rangle$, support an extended trivial
flat region (see Fig.~\ref{fig:fig2_c}(c)).}
Unlike to the earlier case with $\theta_0=0$, we here find that the flatness and the 
non-trivial topology can even co-exist for a single Floquet band; 
$n=2$ band for $\theta_0=\pi/3$ becomes  nearly flat with $C=-1$ around $\theta_1=0.1$ and $1.85$ 
(see Fig.~\ref{fig:fig2_c}(b)). This reflects the fact that 
the relationship between the topology and the flatness, associated
with odd  and even Floquet bands, for $\theta_0=0$, is substantially 
changed for $\theta_0=\pi/3$.

%
Our motivation behind considering two different $\theta_0$ is to selectively tune individual bands as 
topologically flat or trivially flat or topologically dispersive by using another flux parameter 
$\theta_1$. One can obtain the non-topological dispersive bands for
$n=2, 4$ in the whole regime of $\theta_1$  with $\theta_0=0$.
We want to investigate whether one can generate topology and flatness separately  for such bands using other values 
of $\theta_0=\pi/3$. 
Interestingly, we find that, for $\theta_0=\pi/3$, the $n=2$ band can be made topological with $C=-1$ and it
shows non-zero flatness values around $\theta_1=0$ and $\theta_1=1.98$. On the other hand, the $n=3$ band becomes 
topologically trivial, although it shows non-zero flatness for $0\le\theta_1\le2$. 
Therefore, in general, we can dynamically control 
topological and flatness properties of a particular band depending on our requirement. We emphasize that once a given band 
supports topology or flatness or both of them in statics, the Floquet machinery enables us to amplify these initial properties
in a desired manner.

After investigating topology and flatness of energy and quasi-energy bands numerically for the static 
and periodic cases, respectively, we now discuss some interesting aspects of corresponding results. 
One major success of Floquet dynamics is that one can tune the parameters of the system 
such that it contains topologically non-trivial flat bands. Choosing $\theta_0$ in such a way that 
the corresponding static system has topological FBs, we show that the 
Floquet technique allows us to successfully enhance the flux domain within which bands can 
have non-trivial topology and significant flatness compared to the static case. 
{As discussed before, using Floquet dynamics, a particular band can be selectively made trivially flat,  topologically 
dispersive or trivially dispersive.}
On the other hand, the static system does not support any band that shows
non-topological and/or dispersive behavior throughout the whole range of $\theta$. In contrary, Floquet bands
can be made trivial and dispersive irrespective of the value of $\theta_1$.  
Therefore, Floquet driving can indeed pave the way towards a better tunability of the bands by incorporating 
a larger parameter space. 
\textcolor{black}{We note that our aim is to look for the flatness and topology for each individual quasi-energy band. We are not interested in the topological property of a phase as a whole in the driven system.  One can compute $W_3$ invariant in order to properly justify a phase with its corresponding boundary edge modes as discussed above \cite{rudner13}. In order to obtain a complete understanding of the driven system including its individual quasi-energy bands, $W_3$ can be found to be very importanat that we leave for future study (see \cite{supple} for detail).}

\begin{figure} [ht] 
\centering
\includegraphics[width = 3.60cm,height=3.60cm]{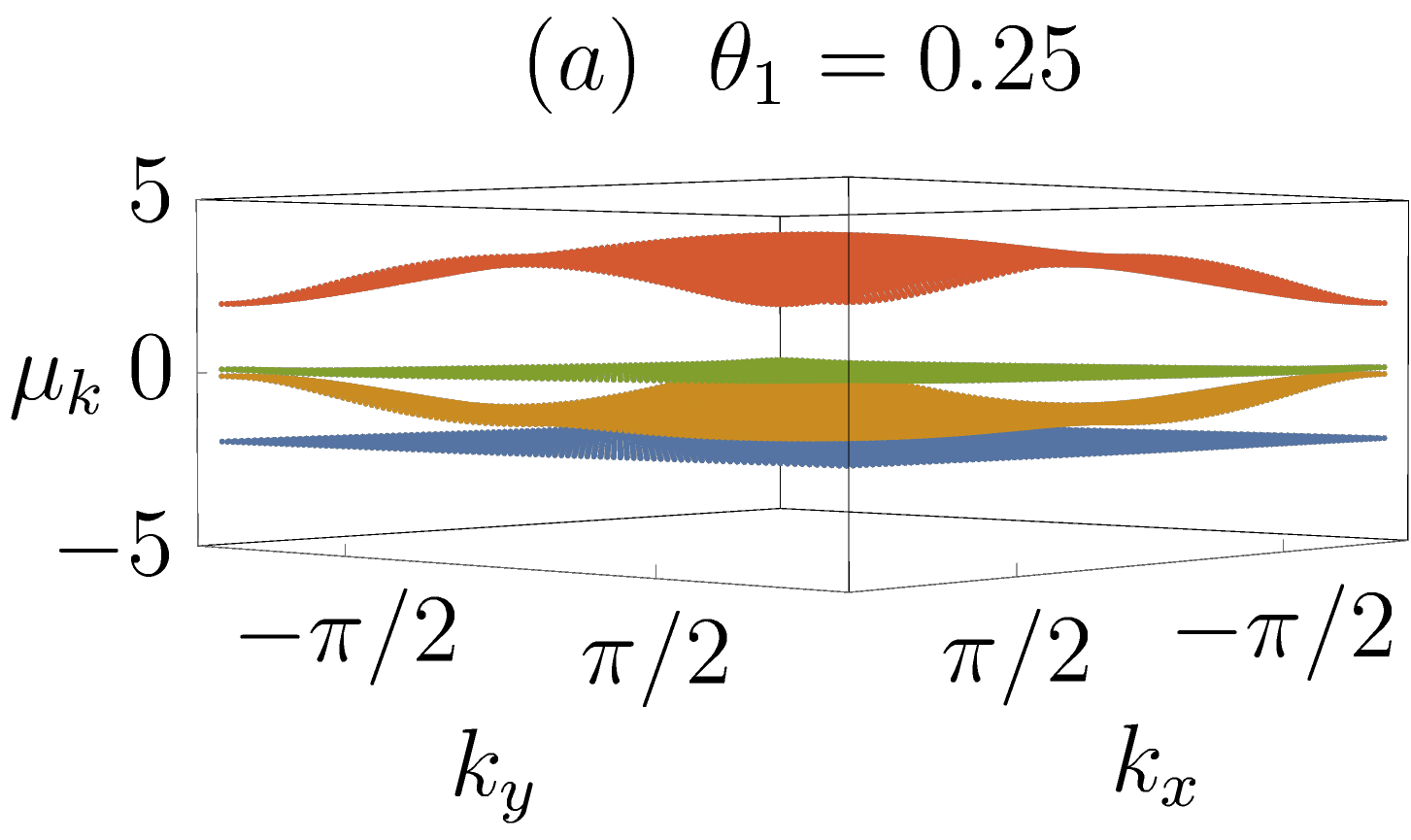}
\includegraphics[width = 3.6cm,height=3.60cm]{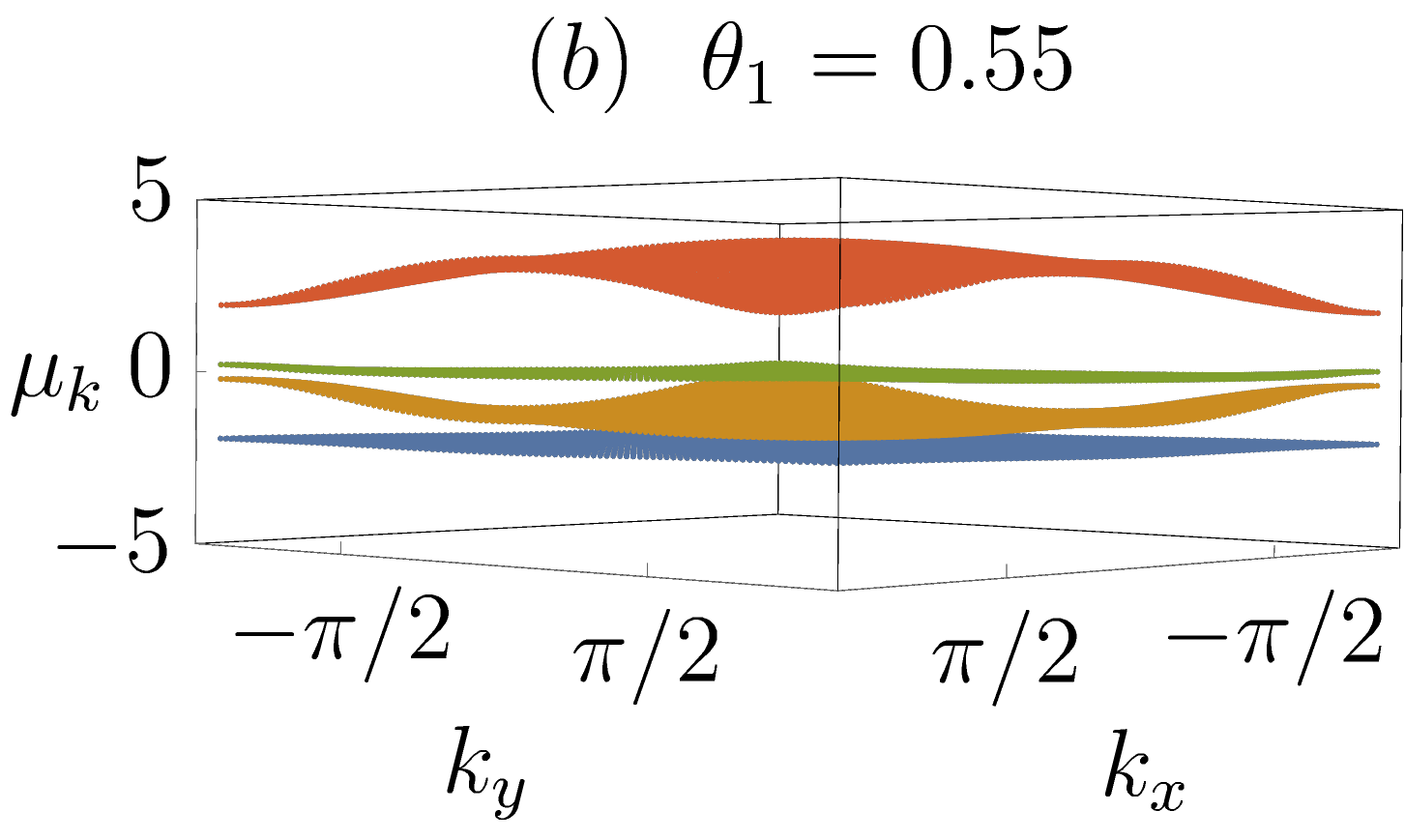}
\includegraphics[width = 3.6cm,height=3.60cm]{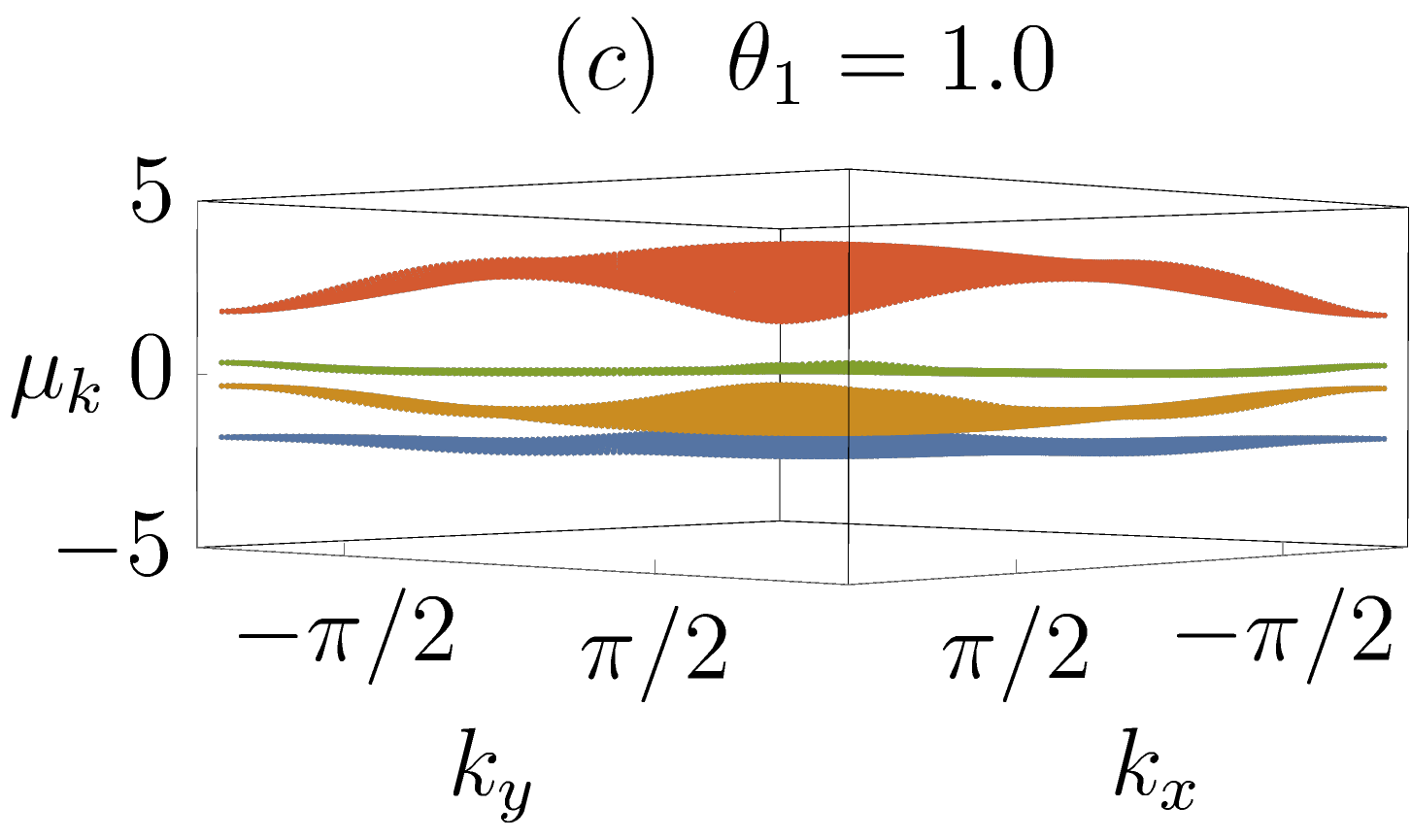}
\includegraphics[width = 3.6cm,height=3.60cm]{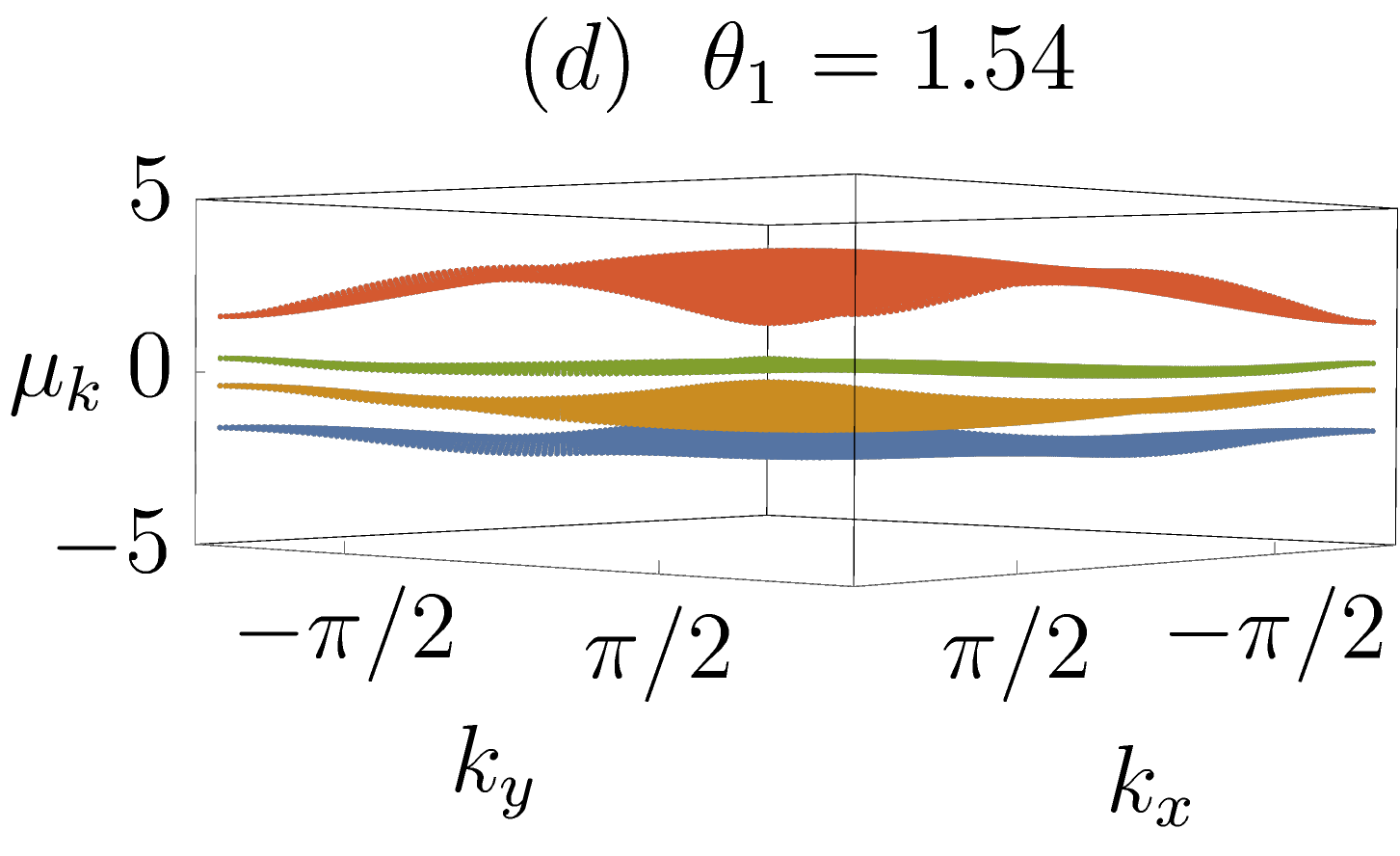}
\caption{Plot of quasi-energy bands in the ${\bm k}$ space for the Floquet operator with the step Hamiltonian 
having $\theta_0=0$ and $\theta_1=0.25$ in (a), $0.55$ in (b), $1.0$ in (c) and $1.54$ in (d). 
It can be observed that the quasi-energy bands $n=1$ and $n=3$ are nearly flat here.
Floquet driving can indeed expand the parameter space in terms of $\theta_1$ for observing flat bands. Here, 
frequency of the driving is considered $\omega=8.0$.}
\label{floquet_band}
\end{figure}

\begin{figure}[ht]
\includegraphics[width=1.0\columnwidth]{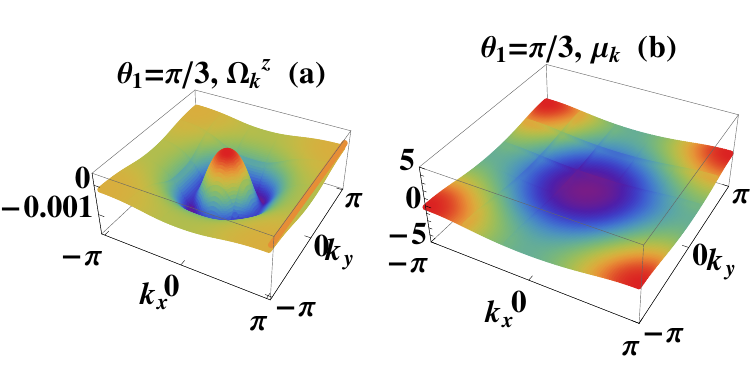}
\caption{(Color online) The Floquet Berry curvature (left panel) and the corresponding ground state ($n=1$) quasi-energy $\mu^{(1)}_{\bm k}$
of the Floquet Hamiltonian (right panel) for $\alpha=0.9$ and $\theta_1=\pi/3$. The quasi-energy band is nearly flat as also 
can be seen from Fig.~\ref{fig:chern_flat_alpha}(b).
We consider $\theta_{\rm ini}=\theta_{0}=0$.}
\label{fig:curv_pib3}
\end{figure}


\begin{figure}[ht]
\includegraphics[width=1.0\columnwidth]{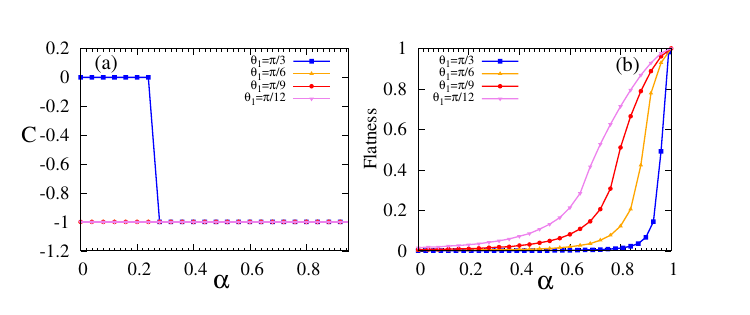}
\caption{(Color online) (a) The Chern number $C$ of Floquet quasi-energy spectrum for $n=1$ is plotted as 
a function of $\alpha$ for different values of $\theta_1$ with $\theta_0=0$. (b) The flatness of the same spectrum 
as a function of $\alpha$ for the same values of $\theta_1$.}
\label{fig:chern_flat_alpha}
\end{figure}

\begin{figure}[ht]
\includegraphics[width=0.8\columnwidth]{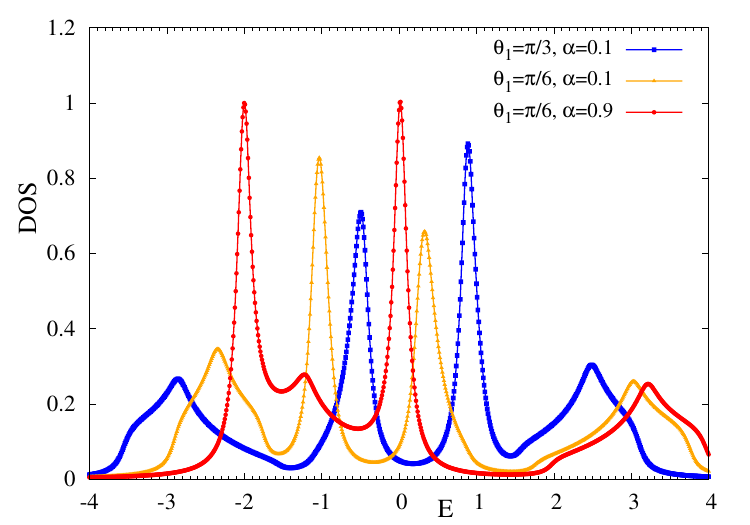}
\caption{(Color online) The Floquet quasi-energy density of states (DOS) for 
different values of $\theta_1$ and $\alpha$. We consider $\theta_{\rm ini}=\theta_{0}=0$.}
\label{fig:dos_floquet}
\end{figure}

\begin{figure}
\includegraphics[width=1.0\columnwidth]{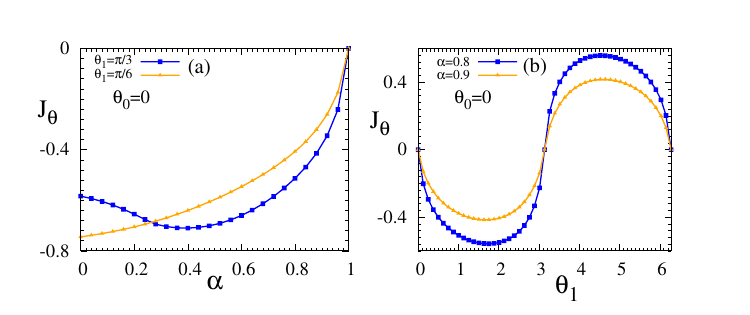}
\caption{(Color online) (a) The total flux current $J_{\theta}$ as a function of $\alpha$ for two values of $\theta_1=\pi/3$ and $\pi/6$ 
after infinite number of step drives $n \to \infty$. (b)
The variation of $J_{\theta}$ with $\theta_1$ for $\alpha=0.8$ and
$0.9$ at $n \to \infty$. We consider $\theta_{\rm ini}=\theta_0=0$.}
\label{fig:jtheta_infinite}
\end{figure}

\begin{figure}
\includegraphics[width=1.0\columnwidth]{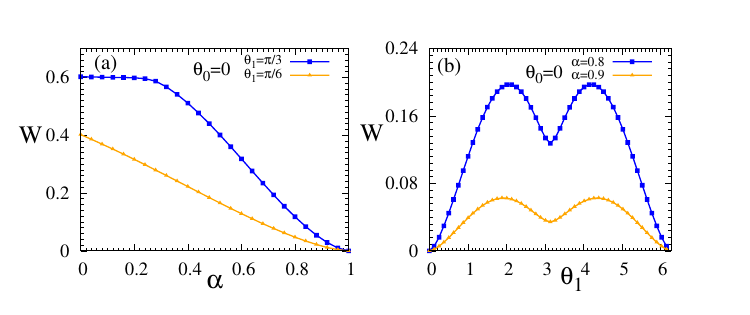}
\caption{(Color online) (a) Plot of the residual energy $W$ as a function of $\alpha$ for $\theta_1=\pi/3$ and $\pi/6$ after infinite 
number of step drives $n \to \infty$. (b) Here $W$ is plotted as a function of $\theta_1$ for $\alpha=0.8$ and $0.9$ at $n \to \infty$.}
\label{fig:w_infinite}
\end{figure}


\textcolor{black}{We now discuss about the Floquet band structures and consider the effect of flux on those.
We also show the distribution of Berry curvature in $(k_x,k_y)$-plane corresponding $n=1$ band under periodic driving. 
The static and Floquet  band structures have been extensively studied in SI \cite{supple}.
In Fig.~\ref{floquet_band}, we have shown Floquet quasi-energy bands in ${\bm k}$-space with $\theta_0=0$ and four 
different values of $\theta_1$. It is found that the bands $n=1$ and $n=3$ are perfectly flat for $\theta_1=0.25$ 
as far as their visibilities are concerned. On the other hand, they become nearly flat as $\theta_1$ is increased further, but the other 
two bands become dispersive. We find that the flatness of $n=1$ and $n=3$ bands sustains up to $\theta_1=1.0$ (with $\theta_0=0$), 
whereas for the static case these bands can only become flat near $\theta \to 0$.
We have already measured the flatness of all the bands for same setting as here 
(see Fig.~\ref{fig:fig2_a}). We can find that our observations for quasi-energy bands show a good agreement with the 
measured flatness of respective bands (see Figs.~\ref{floquet_band} and \ref{fig:fig2_a}).}
In Fig.~\ref{fig:curv_pib3}, we plot the Berry curvature $\Omega_n({\bm k})$ and the Floquet energy $\mu^{(n)}_{\bm k}$ for $n=1$
with $\theta_0=0$ and $\theta_1=\pi/3$.
We can observe that the quasi-energy band is nearly flat for this case.
 It is noteworthy that the distribution of Berry curvature complements the behavior 
of the Floquet band for the above mentioned case.

Having shown the effect of $\theta_0$ and $\theta_1$ on the FBs, 
we next want to investigate the effect of duration of the first step Hamiltonian ${\mathcal H}_{\bm k}(\theta_0)$
in one complete period $T$ by analyzing $C$ and $F$ as a function of $\alpha$ using Eq.~(\ref{eq_FO}).
In particular, the Chern number of $n=1$ band is numerically calculated 
for various  values of $\theta_1$ (see Fig.~\ref{fig:chern_flat_alpha}(a)). We can see that $C$ remains 
at $-1$ for small values of $\theta_1$ such as $\pi/6$, $\pi/9$ and $\pi/12$, during the whole regime 
of $\alpha$. On the other hand, the important observation is that if we start with a comparatively 
larger $\theta_1$ say, $\theta_1=\pi/3$, $C$ remains at $0$ for smaller values of $\alpha$ but 
it becomes $-1$ at $\alpha\approx0.24$ and remains there for further increment of $\alpha$. 
We are now interested 
to find the topologically non-trivial band which are nearly flat. The flatness of the Floquet 
band is numerically calculated as a function of $\alpha$ (see Fig.~\ref{fig:chern_flat_alpha}(b)) 
for the same values of $\theta_1$ as used to determine the Chern number. One can see here 
that flatness of the band is increased with increasing $\alpha$, i.e., increasing the duration 
of $\theta_0$ in Floquet evolution.

The flatness remains at larger values as $\theta_1$ 
is decreased throughout the whole regime of $\alpha$.
This observation is in congruence to Fig.~\ref{fig:fig2_a} for small values of $\theta_1$.
It can be noted that at two extreme values of $\alpha$, i.e., $\alpha=0$ and $1$, 
the values of $C$ and $F$ are solely determined by ${\mathcal H}_{\bm k}(\theta_1)$ and ${\mathcal H}_{\bm k}(\theta_0)$, 
respectively. Therefore, following results at these two extreme values of $\alpha$ are compatible 
with Fig.~\ref{fig:fig1} for the static case. On the other hand, for $0<\alpha<1$, both the Hamiltonians 
are responsible to produce the mentioned results.
The non-zero Chern number for finite $\alpha(<1)$ is an outcome of the Floquet  driving.
We therefore find that one can get a better control to selectively manipulate the 
flatness and topology even by varying $\alpha$ keeping $\theta_1$ fixed. The Floquet operator is a function of 
$\alpha$, $\theta_1$, $\theta_0$; hence, we are able to achieve a large parameter space for 
obtaining topological FBs as compared to the static case which is only restricted to $\theta$.

We here examine another approach to detect the FBs using Floquet 
density of states (FDOS) as $\sum_{{\bm k}, n} \delta(E-\mu^{(n)}_{\bm k})$. However, 
the topological features of the bands can not be probed using this method.
Here we calculate the number of ${\bm k}$ points corresponding to a single quasienergy 
value and plot that number as a function of quasi-energy (see Fig.~\ref{fig:dos_floquet}). 
We find a few peaks in the FDOS for different values of $\theta_1$ and $\alpha$ with $\theta_0=0$. 
If we compare FDOS with the flatness as shown in Fig.~\ref{fig:chern_flat_alpha}(b), 
we can make a connection of the behavior of FDOS with the flatness of the quasienergy  bands.
For $\theta_1=\pi/3$ and $\alpha=0.1$, one can find that the quasienergy band 
($n=1$) is dispersive in nature (see Fig.~\ref{fig:chern_flat_alpha}).
The dispersive nature of the band is clearly reflected in the Fig.~\ref{fig:dos_floquet} where we can see 
that height of the peak around $E=-3$ in the FDOS is much  less than the maximum peak height observed at $E=0,~-2$ for
$\theta_1=\pi/6$ and $\alpha=0.9$. 
For an another case $\theta_1=\pi/6$ and $\alpha=0.1$, the quasienergy band is dispersive (see Fig.~\ref{fig:chern_flat_alpha}(b))
which is again accompanied by a
small broad peak of the DOS around $E=-2.3$ in Fig.~\ref{fig:dos_floquet}. On the other hand, 
for $\theta_1=\pi/6$ and $\alpha=0.9$, we see a large sharp peak in the FDOS around $E=-2$ indicating 
that a FB is supported as the measure of flatness shown in Fig.~\ref{fig:chern_flat_alpha}(b). 
In addition, we observe an identical peak in the FDOS  
at $E=0$ which is corresponding to $n=3$ FB. 
This result confirms the fact that the Floquet flat bands appear in pair for $\theta_0=0$ (see Fig.~\ref{fig:fig1}).

The total flux current (see Eq.~(\ref{jtheta_infty})) and the residual energy (see Eq.~(\ref{eq:re_infinite}))
are also calculated as a function of both $\alpha$ and $\theta_1$ at infinitely large time limit as 
shown in Fig.~\ref{fig:jtheta_infinite} and Fig.~\ref{fig:w_infinite}, respectively. The total current $J_{\theta}$ 
is plotted against $\alpha$ for $\theta_1=\pi/3$ and $\pi/6$. 
{As $\alpha$ increases the total current approaches 
to zero from negative values and finally becomes zero at $\alpha=1$. Since all the four bands contribute in 
calculating total flux current (see Eq.~(\ref{jtheta_infty})), the negative value of $J_{\theta}$ indicates 
that the contribution comes mostly from $n=1$
(\textcolor{black}{See \cite{supple} for detail)}.
The another point to note that, for $\theta_1=\pi/3$, $J_{\theta}$ initially decreases and then starts to increase
following a minimum value around $\alpha=0.4$. This type of behavior is not observed in the case of $\theta_1=\pi/6$, 
where $J_{\theta}$ is a monotonically increasing function of $\alpha$. We can make a connection of such different 
behavior of $J_{\theta}$ for $\theta_1=\pi/3$ with the Chern number of the $n=1$ band for same $\theta_1$. 
It can be seen that the Chern number for $\theta_1=\pi/3$ changes from $0$ to $1$ around $\alpha\approx0.23$, 
whereas it stays at $-1$ for other values of $\theta_1$ (see Fig.~\ref{fig:chern_flat_alpha}(a)).
Although the values of $\alpha$ do not exactly match where the changes occur in $J_{\theta}$ and $C$, we can argue 
that the behavior of $J_{\theta}$ with $\alpha$ has a close connection with the topology of the bands. }

The total flux current also shows periodic behavior with $\theta_1$ as found in each component of the same current 
\textcolor{black}{(see \cite{supple} for detail)}. The point to note here is that $|J_{\theta}(n\to \infty)|$ acquires higher 
value as $\alpha$ decreases i.e., the longer the duration of the second step Hamiltonian ${\mathcal H}_{\bm k}(\theta_1)$:
the flux current of large magnitude is generated in the driven system. Most importantly, the underlying static system, 
as described by ${\mathcal H}_{\bm k}(\theta_{\rm ini}=0)$, does not support flux current, the Floquet dynamics allows 
the system to gain a finite flux current. Similar to each component of the flux current, the total flux current also
crosses zero at $\theta_1=\pi$ where the sign of Chern number also reverses for $n=1$ and $n=3$ bands (see Fig.~\ref{fig:fig2_a}). 
Therefore the total flux current can be used as an indicator of change of topology in the system.
On the other hand, the residual energy of the system decreases with $\alpha$ and vanishes 
at $\alpha=1$. We have also seen that the flatness increases with $\alpha$ for any value of $\theta_1$.
This indicates that the residual energy can be reduced with increasing the flatness 
in the band. We find that the residual energy for $\theta_1=\pi/3$ remains at nearly constant 
value up to $\alpha\approx0.23$ and then monotonically decreases with zero value at $\alpha=1$ 
(see Fig.~\ref{fig:w_infinite}(a)). We have already shown that the Chern number and total 
flux current exhibit distinct behavior for $\theta_1=\pi/3$ as compared to other $\theta_1$'s, this also reflects 
in the behavior of residual energy as a function of $\alpha$. Similar to the flux current, 
$W$ shows oscillatory behavior as a function of $\theta_1$ and the magnitude 
increases with decreasing $\alpha$, as expected. The excess energy can be minimized 
for $\theta_1=\pi$. 
From analysis of different observables as a function of $\alpha$
 and $\theta_1$, we can  convey that our work has experimentally viable, as both of the above 
 parameter can be tuned.
 

\textcolor{black}{We below summarize the main finding of this section. The Floquet FBs (see Fig.~\ref{fig:fig2_a} and  Fig.~\ref{fig:fig2_c}) can be generated and detected using FDOS (see  Fig.~\ref{fig:dos_floquet}).  The static flatness and topology (see Fig.~\ref{fig:fig1}) can thus be tuned with driving parameters $\theta_0$, $\theta_1$ and $\alpha$.  
The demonstration of Floquet quasi-bands are shown explicitely in Fig.~\ref{floquet_band}. The structure of Berry curvature and the topological nature for a given quasi-energy band are shown in  Fig.~\ref{fig:curv_pib3} and  Fig.~\ref{fig:chern_flat_alpha}, respectively.
The evolution of driving induced
flux current $J_{\theta}$ and excess energy $W$ with $\theta_1$ and $\alpha$ are shown in Fig.~\ref{fig:jtheta_infinite} and Fig.~\ref{fig:w_infinite}, respectively. The parameters i.e., duration of the flux Hamiltonian $H(\theta_1)$ and the associated flux $\theta_1$, yield a better control such as, determining the maxima, minima of the above quantities. 
}

\subsection{Aperiodic driving}
\label{apd}

\begin{figure}[ht]
\includegraphics[clip,width=1.0\columnwidth]{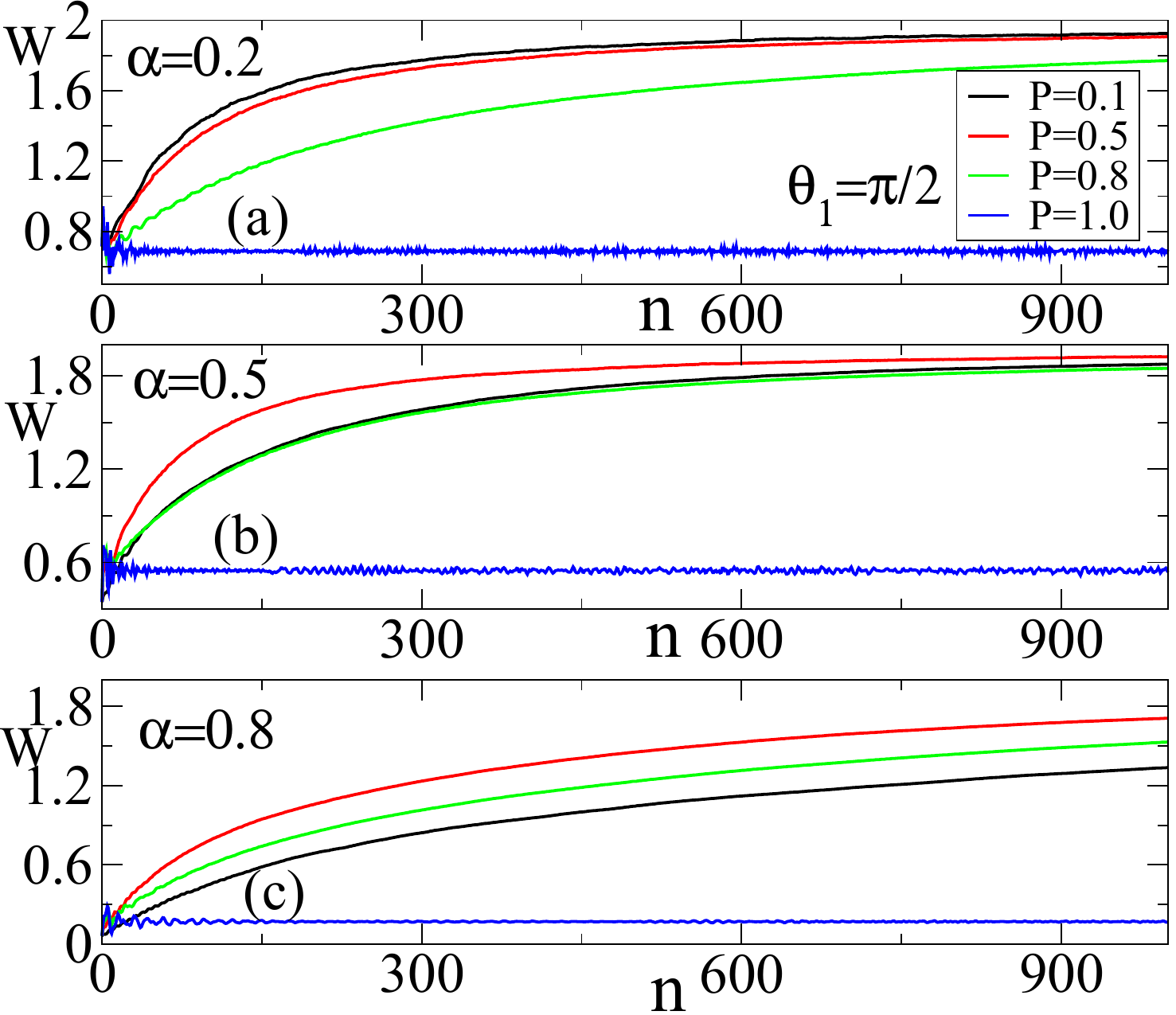}
\caption{Plot depicts the variation of excess energy $W$
as a function of stroboscopic instant $n$ for $\alpha=0.2$ (a),
$\alpha=0.5$ (b), $\alpha=0.8$ (c) with $\theta_1=\pi/2$. 
 For the fully periodic situation ($P=1$), the system synchronizes with the external driving and stops absorbing energy.
 On the contrary, for any non-zero value of P ($\neq 1$), the periodic steady state gets destabilized and the
 system keeps on absorbing heat. For $P=0.5$, $W$ grows maximally
 with $n$ almost independent of the values of $\alpha$. 
 We note that with increasing $\alpha$, the rate of growth of $W$ as a function 
 of $n$ increases.   Here, we have $\theta_{\rm ini}=\theta_{0}=0$.
}
\label{fig:w1_apd}
\end{figure}

After studying the flatness, Chern number, flux current and excess energy in periodic Floquet dynamics, we 
shall now investigate the aperiodic case where $P$ denotes the probability of appearing the second flux Hamiltonian 
in the driving protocol as defined in Eq.~(\ref{eq:3}).
We shall first study the dynamics of excess energy $W$, as obtained from Eq.~(\ref{eq_rs1}) and Eq.~(\ref{eq:re}), by varying 
$\theta_1$ and $\alpha$. Figure \ref{fig:w1_apd} shows that 
for $\theta_1=\pi/2$ and $\theta_{\rm ini}=\theta_0=0$, $W$ increases less rapidly for $\alpha=0.8$.
\textcolor{black}{A close observation of the numerical results for different $\alpha$ indicates that $W$ stays at maximum 
value as a function of time for $P=0.5$, compared to any $P \neq 0.5$ for $\alpha>0.2$.}
This implies the fact that the system absorbs energy maximally 
when the degree of  aperiodicity is maximum, while for 
fully periodic drive $P=1$, the system attains a periodic steady state i.e., 
it does not absorb energy from the external drive. Interestingly, 
 any amount of aperiodicity can drive the system
 away from the periodic steady state and hence it
 gets heated up with time. While investigating with $\alpha$, for a given value of $n<300$,
 we find that short duration of flux-Hamiltonian (i.e., $\alpha=0.8$) can lead to the decrement of $W$ as compared to the long
 duration of flux Hamiltonian (i.e., $\alpha=0.2$).
 The finite time rate of growth of $W$ becomes higher for $\alpha=0.2$ while the intermediate rate of 
 growth becomes higher for $\alpha=0.8$. As a result,  the asymptotic value of $W$ is reached early for $\alpha=0.2$ while 
 the $W$ saturates for much higher value of $n$ for $\alpha=0.8$.  We note that the 
 asymptotic value depends on $\alpha$.  The reason behind the above characteristic will be discussed below. 
\textcolor{black}{The point to note here is that the heating in the system gets remarkably 
suppressed as we reduce the value of $\theta_1$ for any $\alpha$ (See \cite{supple} for detail). }

We would now try to understand the physical picture behind the rise of $W$ and its subsequent saturation 
at $n\to \infty$. It has been shown for a two level model (in the momentum space) that aperiodic dynamics can be analytically handled 
in a non-perturbative way \cite{bhattacharya18,maity18}; the instantaneous energy $e_{\bm k}(nT)$ as defined 
in Eq.~(\ref{eq_rs1}) is proportional to 
$(D_{\bm k})^n$ while the proportionality factor depends on initial state $|\Psi_{\bm k}(\theta_{\rm ini},t=0)\rangle$ and 
possible combinations of Floquet basis. The disorder matrix $D_{\bm k}$ depends on  $P$, $T$, Floquet operator ${\mathcal F}_{\bm k}$ and
eigen-energies of ${\mathcal H}_{\bm k}(\theta_0)$. 
For two level system with binary disorder ($g_n$ can be 
either $1$ or $0$), $4\times 4$ disorder matrix has two real (unity and less than unity)  and two complex conjugate 
(magnitude less than unity) eigenvalues.
Now for small $n$, the complex eigenvalues dictate the oscillatory pattern on the 
overall increasing background. The increasing nature, persisted until $n$ becomes substantially large, is dictated 
by the real eigenvalue which is less than unity . The asymptotic universal nature is solely determined by the unity eigenvalue as all the other contributions
coming from the remaining eigenvalues vanish.  Therefore, it is understood that 
$W$ can not be raised indefinitely rather than it saturates as $n \to \infty$. These saturation values depend on the 
proportionality factors. Now connecting it to our case, one can similarly construct 
a $8\times 8$ disorder matrix $D_{\bm k}$ as our momentum space Hamiltonian is $4\times 4$. We will be having 
unity eigenvalue in $D_{\bm k}$ that determines the asymptotic results at $n\to \infty$. All the 
remaining eigenvalues dictates the low and intermediate growth of $W$.

Turning to periodic dynamics with 
$P=1$, we observe by analyzing Fig.~\ref{fig:w1_apd} that the system attains periodic steady state 
with minimum excess energy when the flux Hamiltonian is activated for 
short duration of time within one time-period and flux $\theta_1 \to 0$. Both of the 
above observations are associated with the fact that the initial state is the 
ground state of Hamiltonian with $\theta_0=0$. For aperiodic case, 
we also note that the asymptotic value of $W$ at $n\to \infty$ is a function of 
$\theta_1$ and $\alpha$.

\begin{figure}[ht]
\includegraphics[clip,width=1.0\columnwidth]{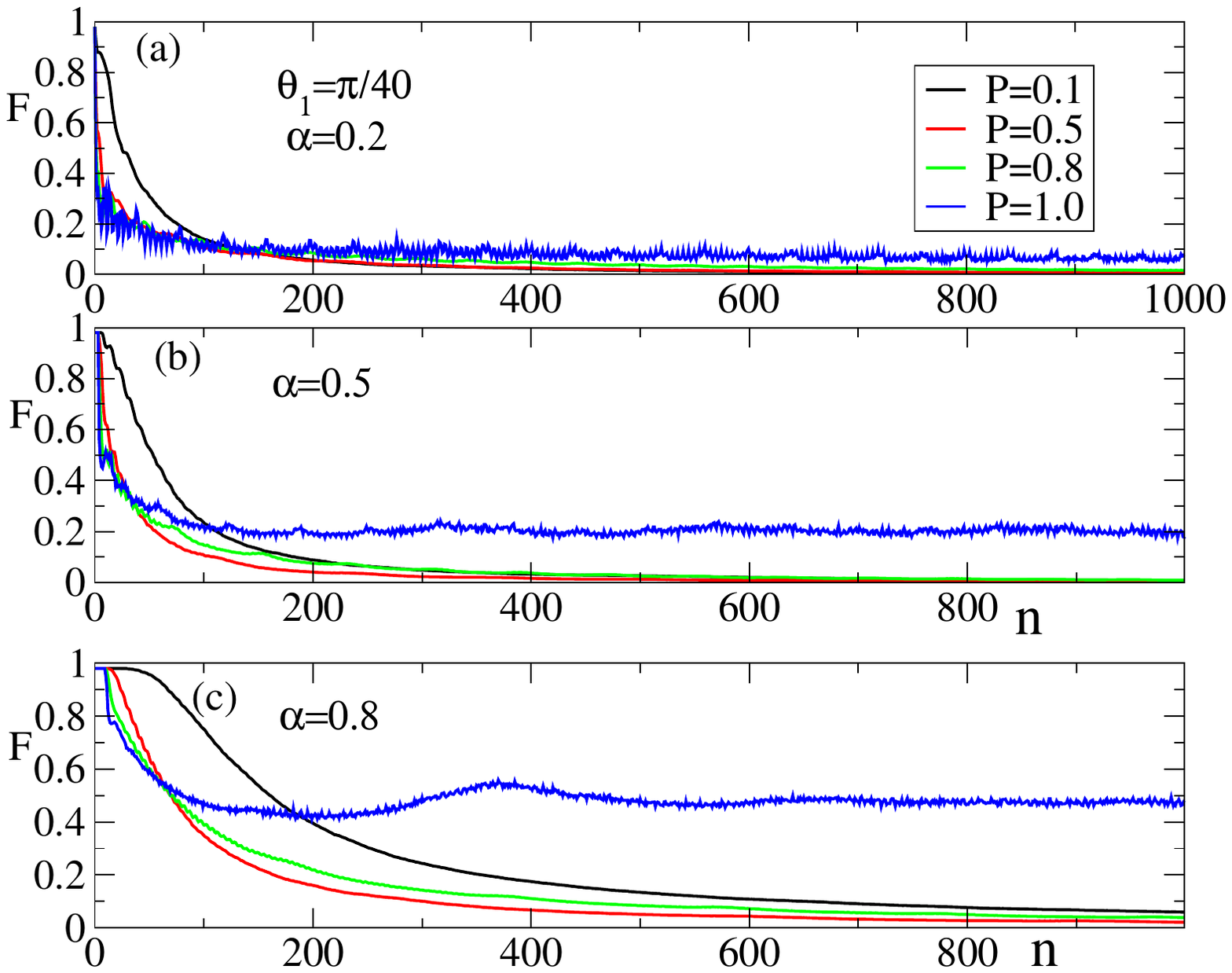}
\caption{ Plot depicts the variation of
instantaneous flatness $F$ 
as a function of stroboscopic instant $n$ for $\alpha=0.2$ (a),
$\alpha=0.5$ (b), $\alpha=0.8$ (c) with $\theta_1=\pi/40$.
One can see that periodic and aperiodic driving both are not able 
to generate flat band. We here choose $\theta_{\rm ini}=\theta_{0}=0$.
The instantaneous flatness acquires higher values for periodic dynamics.
We here choose $\theta_{\rm ini}=\theta_{0}=0$.
}
\label{fig:f3_apd}
\end{figure}

We shall now study the dynamical flatness by varying different parameters such as $\theta_1$, 
$\alpha$ and $P$. 
We show that for $\theta_1\gg 0$, the time evolved instantaneous energy becomes 
highly dispersive irrespective of the duration $\alpha T$ of the no-flux Hamiltonian. 
This has been quantified by calculating $F$ using Eq.~(\ref{eq:flatness_2}) and
shown in Fig.~\ref{fig:f3_apd}. 
 The time evolved $e_{\bm k}(nT)$ can have non-dispersive nature as far as 
 small $n$ is concerned. The aperiodic 
 driving  leads to significant flatness as compared to the 
 periodic driving for small but finite $n$ limit. 
 \textcolor{black}{This scenario is clearly visible for small $\theta_1$, say $\theta_1=\pi/40$, 
 irrespective of the duration of the no-flux Hamiltonian
 (see Fig.~\ref{fig:f3_apd})}
 We find that for  periodic driving with $P=1$,
 $e_{\bm k}(nT)$ becomes less dispersive if one increases 
 the duration of no-flux Hamiltonian from $\alpha=0.2$ to $0.8$. 
 We see that $F$ remains at closely unit value for $P=0.1$ with 
 $\alpha=0.8$, while flatness falls most rapidly for 
 $P=1$ as far as $n \le 200$. This is due to the fact that the
 time-evolved wavefunction $\lvert \Psi_{\bm k}(nT)\rangle$ is 
 minimally deviated from the initial wave-function for 
 $P=0.1$ and $\alpha=0.8$, i.e., the system is closely 
following trivial evolution as the flux-Hamiltonian  is mostly inactive and 
its duration is very short over a time-period. At the same time, we know
that the initial no-flux Hamiltonian supports non-topological flat band as the ground state. 
Hence, for such aperiodic driving, the flatness remains nearly constant at unity for 
small time and then starts decaying since flux-Hamiltonian even for short duration is effectively 
active at large time. On the other hand, for periodic case with $P=1$, due to the similar reason 
the fluctuation of $F$ gets reduced heavily when $\theta_1 \to 0$ and $\alpha\to 1$. \textcolor{black}{ The flatness of the instantaneous state $| \Psi(t=nT) \rangle$ 
is closely connected to the 
the survival probability of an initial state $P_s= |\langle \Psi(t=0,\theta_{\rm ini})| \Psi(t=nT)\rangle|^2 $. From the behavior of instantaneous flatness it can be estimated that $P_s$ decays with $n$, however, there  always exists a finite $P_s$ even for large $n$ (See \cite{supple} for detail).}

\begin{figure}[ht]
\includegraphics[clip,width=1.0\columnwidth]{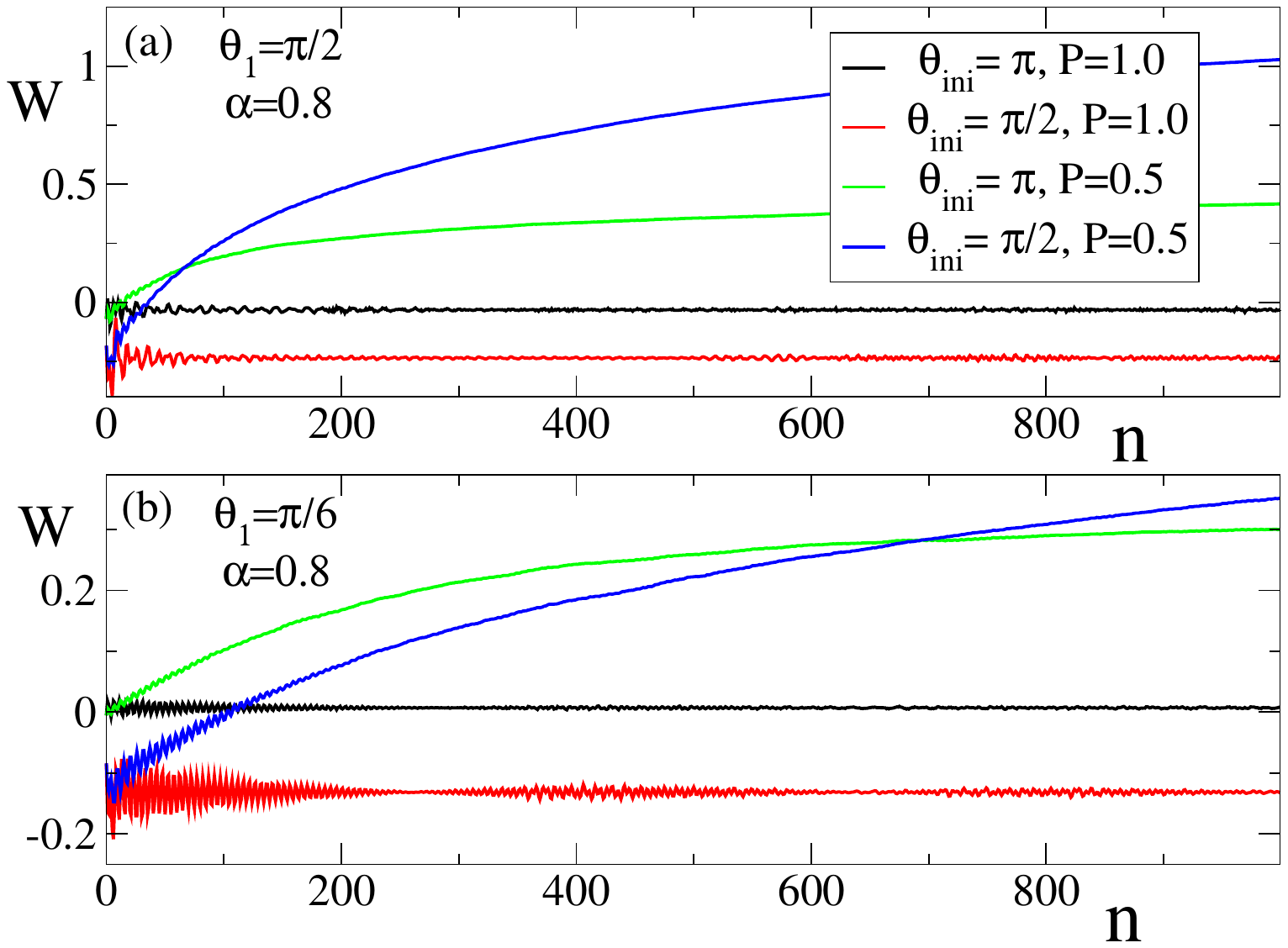}
\caption{Plot depicts  the variation of
instantaneous work with stroboscopic cycle $n$ with two step Hamiltonian
${\mathcal{H}}_{\bm k}(\theta_0=0)$ and ${\mathcal{H}}_{\bm k}(\theta_1=\pi/2)$
considering the ground state of ${\mathcal{H}}_{\bm k}(\theta_{\rm ini})$
as initial state $|\Psi_{\bm k}(\theta_{\rm ini,t=0})\rangle$. (a) for 
$\theta_1=\pi/2$ and (b) for $\theta_1=\pi/6$. We find for the 
aperiodic driving with $P=0.5$, $W$
 increases more rapidly once we start from dispersive band at $\theta_{\rm ini}=\pi/2$
 compared to a FB at $\theta_{\rm ini}=\pi$. However, for periodic driving 
 $P=1$, $W$ saturates at higher value starting from dispersive band compared 
 to FB. Here, $\alpha=0.8$.}
\label{fig:w1_apd_pd}
\end{figure}

\begin{figure}[ht]
\includegraphics[clip,width=1.0\columnwidth]{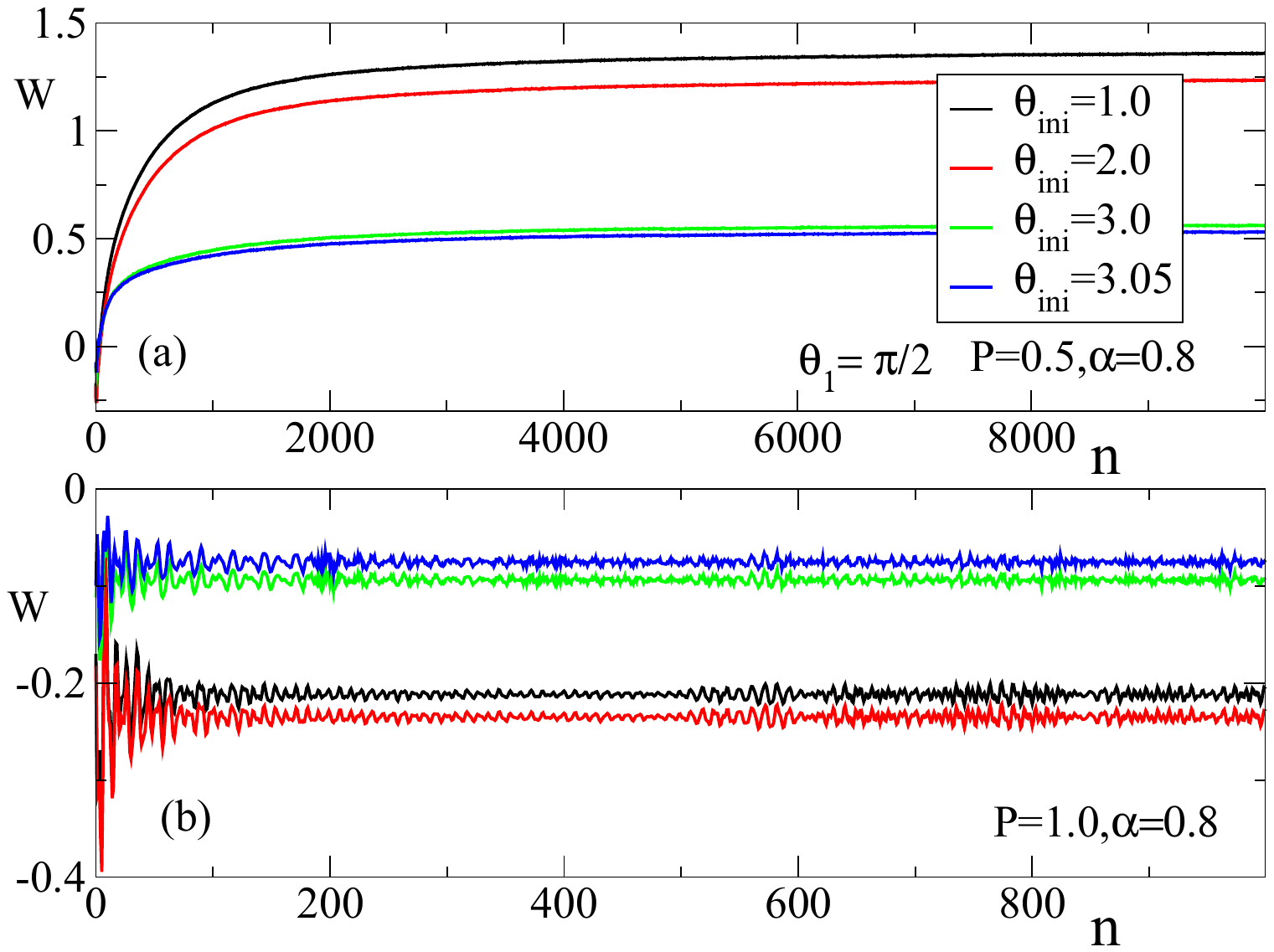}
\caption{Plot depicts  the variation of
instantaneous work done as a function of stroboscopic cycle $n$ for 
aperiodic $P=0.5$ (a) and periodic $P=1.0$ (b) driving with $\theta_1=\pi/2$. 
Starting from a FB ($\theta_{\rm ini}=3.0,~3.05$), $W$ saturates to a lower value as compared to 
dispersive band ($\theta_{\rm ini}=1.0,~2.0$). Here,  $\alpha=0.8$.
}
\label{fig:w2_apd_pd}
\end{figure}

Until now we have considered the situations with $\theta_{\rm ini}=\theta_{0}$, we will
now investigate the case where  $\theta_{\rm ini} \ne \theta_{0}$ 
i.e., initial state is not the eigenstate of the first step Hamiltonian 
${\mathcal H}_{\bm k}(\theta_{\rm ini})$.
Considering two different $\theta_{\rm ini}$ and keeping $\theta_1$, $\theta_0$
fixed, we will be   able to compare the two equivalent non-eigenstate evolution 
as far as the first step Hamiltonian ${\mathcal{H}}_{\bm k}(\theta_0)$ is 
concerned. Using this method, we can analyze the effect of initial flat 
as well as dispersive bands on $W$ following the identical dynamical protocol. 
We investigate the residual or excess energy $W$ as a function of 
$n$ considering different initial states and a variety of step Hamiltonians for 
the periodic and aperiodic driving as shown in Fig.~(\ref{fig:w1_apd_pd}), 
Fig.~(\ref{fig:w2_apd_pd}). The interesting 
outcome is that in the case of aperiodic driving the growth rate
of $W$ becomes heavily slowed down if the initial state has a significant 
flatness.

We study the evolution of excess work $W$ for $\theta_1=\pi/2$ and $\pi/6$
with $P=1$ and $0.5$ as shown in  Fig.~\ref{fig:w1_apd_pd}(a) and (b), respectively. 
The excess energy $W$ turns out to be negative for the periodic driving 
and also for the aperiodic driving for small time; this is due to the fact that 
instantaneous energy $e_{\bm k}(nT)$ is less than the initial 
energy $e^{\rm ini}_{\bm k}(0)$ (see Eq.~(\ref{eq:re})). In the present case, $e^{\rm ini}_{\bm k}(0)$ is
not an eigenstate energy. One can observe that the  periodic driving is not able to excite the system. 
{However, aperiodic driving can not lead to a periodic steady state
and hence, instantaneous energy $e_{\bm k}(nT)$ can overcome  $e^{\rm ini}_{\bm k}(0)$.}
Most interestingly, the growth rate of $W$ is significantly suppressed 
once the initial state has substantial flatness; the energy eigenvalues corresponding to the state
$|\Psi_{\bm k}(\theta_{\rm ini},t=0)\rangle$
are more flat for $\theta_{\rm ini}=\pi$ as compared to $\pi/2$. In order to 
examine these features in detail, we repeat the aperiodic case with $P=0.5$
as depicted in Fig.~\ref{fig:w2_apd_pd}(a) for $\theta_1=\pi/2$. One can clearly observe that $W$ calculated from initial 
dispersive band saturates at a higher value compared to the initial FB. On the other hand,
for the periodic case, the behavior is the opposite (see Fig.~\ref{fig:w2_apd_pd}(b)); 
$W$ obtained from an initial dispersive band saturates at a smaller value as compared to 
an initial FB. The reason is $e^{\rm ini}_{\bm k}(0)$ for dispersive band becomes lower compared to the FB. \textcolor{black}{This behavior as expected does not change for aperiodic dynamics.}
Since the system does not keep absorbing energy from the periodic driving, $W$ for dispersive band always 
stays lower compared to that of the FB.

We can clearly observe that the initial flatness of a band has a significant effect 
on the subsequent dynamics. For the aperiodic case, as we know that the
disorder matrix $D_{\bm k}$ does not depend on  initial condition, rather it is the 
proportionality factor that depends on the initial condition. In the present 
analysis by keeping the two step flux Hamiltonians fixed, we consider different 
situations by varying initial conditions. Therefore, we actually change the proportionality factor 
instead of changing the disorder matrix.
\textcolor{black}{One can find that both the instantaneous energies $e(\theta_{\rm ini}=3.05,~3.0,t=nT)$ and $e(\theta_{\rm ini}=1.0,~2.0,t=nT)$ saturate to an identical value irrespective of the initial flatness as observed in terms of the survival probability $P_s$ (see Sec.~VII of the Ref.~\cite{supple} for more detail). The  instantaneous energy  rises more with time and eventually leading to a longer  saturation time while starting from an initial FB rather than a non-flat band. }
This may be the reason to saturate the excess energy $W$ 
to a higher positive value when the initial state is substantially flat. This initial 
state dependence is further confirmed by varying
the dynamical parameter $P$ and $\theta_1$ while keeping $\theta_{\rm ini}$
fixed (see Fig.~\ref{fig:w2_apd_pd}). Moreover, 
the small time rise of $W$ is almost identical for the different 
$\theta_{\rm ini}$ as discussed above. Similar to the 
asymptotic value, the intermediate growth rate of $W$ increases 
for initial dispersive band as compared to the FBs.

\textcolor{black}{We below provide a plausible argument behind above oberservation.
The high degeneracy of the initial FB 
can act as an energy absorbing agent while the system is driven
out of equilibrium. The excited quasi-particle due to driving can not fill the states above until they occupy all the
degenerate states of the FBs. The quasi-particles associated with a flat band have the same energy. To excite
those at the higher level, each quasi-particle needs same amount of energy from the driving. As a result, system needs
a large amount of energy to excite all the quasi-particles. In comparison, the initial dispersive band hosts non-degenerate quasi-particles that are easily excited by absorbing energy from driving. As a result, the subsequent dynamics starting from initial FB and dispersive band are quite different. }

\textcolor{black}{We demonstrate the evolution of excess energy and instantaneous flatness in Fig.~\ref{fig:w1_apd} and Fig.~\ref{fig:f3_apd}, respectively. We here find that aperiodic (periodic) driving leads to heating (non-equilibrium steady state without heating) and consequently a substantial dispersiveness is generated in the time evolved state. Now the heating can be reduced while we start from an initial flat band rather than dispersive band as shown in Fig.~\ref{fig:w1_apd_pd}. We next verify  this finding by considering different combination of initial states and Floquet operators as shown in Fig.~\ref{fig:w2_apd_pd}. }

\section{Experimental feasibility}
\label{expt}

\textcolor{black}{From previous literatures on the flat bands, we know that they arise due to the destructive 
quantum phase interference of fermion hopping paths in tight-binding systems on different lattices. There 
could be two possible experimental techniques to fabricate such lattice structure in laboratory; one is 
photonic waveguide and the other is optical lattice. The aberration-corrected femtosecond laser-writing method
can efficiently yield a precise fabrication of two-dimensional arrays of sufficiently deep single-mode
waveguides. The two-dimensional Lieb lattice structure~\cite{Vicencio15,Mukherjee15} and other lattice 
geometries~\cite{Longhi14,Zong16} are successfully realized using photonic waveguides. 
The diamond-octagon lattice structure can also be realized experimentally using photonic waveguide 
with the values of the experimental parameters as: the lattice period $20-30$ ~$\mu$m, 
propagation distance $ 7-10$ cm, and operating wavelength $500-800$ nm~\cite{Vicencio15,Mukherjee15_2}. 
The topological properties of a lattice structure can also be investigated experimentally using optical 
waveguide~\cite{Zhong19}. On the other hand, ultracold atomic condensates in optical lattices can artificially 
generate lattice structure by suitably tuning the hopping amplitudes, interaction strength and potential depth~\cite{bloch-rmp2008}.
The tunable effective magnetic field can be generated for ultracold atoms in optical lattices~\cite{Aidelsburger11}.
Similarly, we can realize our model in the laboratory using ultracold atoms in optical lattice. To observe our results, 
the frequency of the step driving has to be large as compared to the band-width of the system and the magnetic field 
has to be comparable with the square of the lattice spacing such that $\theta$ becomes finite $[0,2 \pi]$. As far as 
the numerical values of the experimental parameters are concerned, we estimate those as,
switching frequency $\omega \sim 1-100$ ev with $t_x,~t_y,~\lambda \sim 0.1-10$ ev such that $\omega >t_x,~t_y,~\lambda$, 
switching time $T \sim 0.01-1.0$ fs and the magnetic field $B \sim 0.1-10.0 $ nT. Moreover, in addition to the  optical lattice platform we hope that our 
result can be tested in various metamaterials such as photonic \cite{exp1,exp2,exp3}, acoustic \cite{exp4,exp5} lattices and solid state systems \cite{exp6}.}

\section{Conclusion}
\label{summary}
Motivated by the recent equilibrium studies on topological FBs \cite{Pal18}, we here investigate a two dimensional diamond-octagon model 
with time dependent flux Hamiltonian. To be precise, the driving protocol considered here is periodic and each cycle is
comprised of two step Hamiltonians corresponding to two different magnetic fluxes $\theta_0$ and $\theta_1$ embedded in them. 
This Floquet set up allows us to characterize the driven model in terms of the two parameters, 1) duration of first flux Hamiltonian 
$\alpha T$ and 2) values of flux $\theta_i$. Most interestingly, we show that
the topological FBs can be engineered quite desirably by appropriately tuning these above parameters for which the
static model does not support topology and FB simultaneously. In the process, we  provide a new
definition of flatness and check the
consistency of our definition using Floquet density of states (FDOS) where we find sharp peak at the FB energy.
We calculate Chern number $C$ and flatness of the Floquet quasi-energy bands by varying $\alpha$ and $\theta_1$ 
to get Floquet topological flat bands. We also show the emergence of flux current corresponding to each Floquet band 
and describe how this is connected to change of topology in the system.
Considering the asymptotic limit of the number of driving cycle $n \to \infty$,
we additionally study the total flux current $J_{\theta}$ and the variation of excess energy $W$ 
as a function of $\alpha$ and $\theta_i$.
Interestingly, $J_{\theta}$ and $W$ both show periodic behavior with $\theta_1$. We also show the interconnections between topology of the bands, flux current, flatness and 
excess energy of the driven system. Importantly, we show that the excess energy due to periodic drive in the system can be reduced 
by increasing the initial flatness of the energy bands.

We next analyze the stroboscopic evolution of $W$ and flatness with $n$  considering aperiodic driving.
Here the protocol we follow is that the second Hamiltonian inside the cycle is associated with a binary disorder amplitude with probability $P$.
 In this way, we can go to perfect periodic limit
for probability $P=1$ and, $P=0$ corresponds to a situation where
the system is evolved only with the first Hamiltonian. Any intermediate 
value of $P$ corresponds to a random array of these two Hamiltonian. We show that
$W$ can be substantially suppressed if $\alpha \to 1$ and $\theta_1 \to 0$. 
On the other hand, maximum heating occurs when $\alpha \to 0$ and $\theta_1 \to \pi/2$. 
For periodic case, $W$ saturates to a higher value for $\alpha \to 0$ and $\theta_1 \to \pi/2$ 
compared to $\alpha \to 1$ and $\theta_1 \to 0$. We also study the instantaneous  flatness that can only sustain with 
periodic dynamics. Most interestingly,  
starting from an initial FB, $W$ saturates at a lower value for the aperiodic case;
a large number of  degenerate states associated with the FBs
are responsible for this suppression. We further explain our observation by making resort to the disorder matrix representation of the work done. 

{In the case of aperiodic driving, we mainly ask the questions that how we can minimize or suppress heating in the system. 
In experiments, realizing a purely periodic drive is a very difficult task. Therefore, aperiodic drive is more realistic case than the 
periodic one. On the other hand, heating is a real problem in such cases to realize interesting Floquet phases such as Floquet topological 
flat bands in the present set up. Interestingly, we show that the heating can be suppressed by reducing the value of flux parameter 
$\theta_1$ in the case of aperiodic driving. Further it can be minimized by choosing appropriate initial state that supports non-dispersive 
Floquet bands. In this context, our work seems to be useful for its practical implications.}

\textcolor{black}{Interestingly, the effect of interactions for  
systems with flat bands, showing high density of states, is an active area of research \cite{wu-prl2007,wang13}. The interaction mediated strongly correlated phenomenon such as, superconductivity and fractional quantum Hall effect, thus can come into play when the 
degeneracy is
lifted by the interaction in these systems. It would be indeed interesting to come up with a more realistic model considering the  electron-electron Hubbard interaction as the future study. At the same time,  delocalization-localization transitions for many-body systems receive enormous attention \cite{lazarides15}. Ours study on dynamics of excess energy  can be further extended for interacting system that is beyond the scope of the present study. Being focused on the non-interacting systems, one possible future direction could be to explore the dynamic winding number $W_3$ invariant extensively for the flat bands while the present study is only limited to Chern number.}



 
\appendix

 

\end{document}